\documentclass[aps,prl,twocolumn,superscriptaddress,english]{revtex4}
\usepackage{amssymb}
\usepackage{graphicx}
\usepackage{amsmath}
\usepackage{soul}
\usepackage{amsthm}
\usepackage{amsfonts}
\usepackage[T1]{fontenc}
\usepackage[latin9]{inputenc}
\usepackage{array}
\usepackage{multirow}
\usepackage{color}
\usepackage{esint}
\usepackage{bm}
\usepackage{color}
\usepackage{bbm}
\usepackage{hyperref}
\usepackage{babel}
\usepackage{titlesec}

\begin{document}

\global\long\def\id{\mathbbm{1}}
\global\long\def\ui{\mathbbm{i}}
\global\long\def\ud{\mathrm{d}}

%\title{Extending the concept of mobility edges; generalized}
\title{Extended Landauer-B\"{u}ttiker Formula for Current through Open Quantum Systems with Gain or Loss}
\author{Chao Yang}
\affiliation{Shenzhen Institute for Quantum Science and Engineering,
Southern University of Science and Technology, Shenzhen 518055, China}
\affiliation{International Quantum Academy, Shenzhen 518048, China}
\affiliation{Guangdong Provincial Key Laboratory of Quantum Science and Engineering, Southern University of Science and Technology, Shenzhen 518055, China}
\author{Yucheng Wang}
\thanks{Corresponding author: wangyc3@sustech.edu.cn}
\affiliation{Shenzhen Institute for Quantum Science and Engineering,
	Southern University of Science and Technology, Shenzhen 518055, China}
\affiliation{International Quantum Academy, Shenzhen 518048, China}
\affiliation{Guangdong Provincial Key Laboratory of Quantum Science and Engineering, Southern University of Science and Technology, Shenzhen 518055, China}
\begin{abstract}
	The Landauer-B\"{u}ttiker formula, which characterizes current through a finite region connected to leads, is a cornerstone in studying transport phenomena. We extend this formula using the Lindblad-Keldysh formalism to describe particle and energy currents in regions with gain or loss. The derived formula shows that gain or loss plays a notable role in shaping the transport properties. Based on this formula, we find that inversion symmetry breaking in the gain and loss terms or in the system can generate current, along with the anomalous phenomenon where disorder induces current. We also reveal how gain and loss affect thermal and electrical conductances, as well as the impact of bond loss-induced non-Hermitian skin effect on current. This work deepens and extends our understanding of transport phenomena in open systems.
	
\end{abstract}
\maketitle

{\em Introduction.---} Studying quantum transport properties is essential for understanding condensed-matter systems and has broad applications, such as in various electronic devices, atomic and photonic technologies, nanotechnology, and quantum computing~\cite{review0,review1,review2,review3,review4,review5,review6}. 
One of the most influential frameworks for understanding transport is the Landauer-B\"{u}ttiker formula~\cite{LB1,LB2}, which expresses current in terms of the system's transmission coefficient and the electron distribution within connected leads. Beyond electrons, this formalism applies universally to any quantum system--including atomic~\cite{BLtransport1,BLtransport2,BLtransport3,BLtransport4,BLtransport5}, photonic~\cite{BLtransport6,BLtransport7}, and phononic~\cite{BLtransport8} systems--where transport reduces to a scattering problem with well-defined transmission channels~\cite{BLnote}. 
The system described by this formula, in its original coherent formulation, is not influenced by any form of dissipation other than the coupling to the two leads~\cite{explainOLB,ButtikerVP}.
However, dissipation is widespread, and understanding its effects on quantum transport is a fundamentally important issue.

With the advancements in experimental techniques for controlling different types of dissipation~\cite{technology0,technology1,technology2,technology3,technology4,technology5,technology6,technology7,technology8,technology9,technology10,technology11,technology12,technology13,technology14,technology15,technology15s1,technology15s2,technology15s3,technology16,technology17,technology18,technology19,technology20,technology21,technology21s,technology22,technology23,technology24,technology25,technology26,technology27}, recent years have seen a growing interest in dissipative open quantum systems. While dissipation is typically viewed as detrimental to quantum correlations, recent research has revealed that it can also give rise to novel physical phenomena~\cite{technology3,technology4,technology5,technology6,technology7,technology8,technology9,technology10,technology11,technology12,technology13,technology14,technology15,technology15s1,technology15s2,technology15s3,technology16,technology20,technology21,technology21s,technology22,technology23,technology24,technology25,technology26,technology27,phenoma1,phenoma2,phenoma3,phenoma4,phenoma5,phenoma6} or phase transitions~\cite{PT1,PT2,PT3,PT4,PT5,PT6,PT7,PT8,PT9,PT10,PT11,PT12}, profoundly impacting dynamics. A common type of dissipation is dephasing, which can reduce coherence, thereby breaking localization and ballistic transport, and inducing diffusive transport~\cite{Loc1,Loc2,Loc3,Loc4,Loc5,Loc6,Loc7,Loc8,Loc9,Loc10,Loc11,Loc12,Loc13}. 
Another important type of dissipation is particle gain or loss. 
This process can arise from a variety of physical mechanisms, such as proximity effects in heterostructures~\cite{PE1,PE2,PE3,PE4}, uncontrolled leakage in experimental setups~\cite{UL1,UL2,UL3}, or engineered coupling in platforms like cold atoms, photonic lattices, superconducting circuits, and ion traps~\cite{technology13,technology14,technology15,technology15s1,technology15s2,technology15s3,technology16,technology17,technology18,technology19,technology20,technology21,technology21s,technology22,technology23,technology24,technology25,technology26,technology27}. %When the environment satisfies the Markovian condition, its effect can be effectively captured by gain/loss terms in the Lindblad master equation. 
Despite the increasing relevance of the transport properties in such systems ~\cite{technology16,technology17,technology18,technology19,technology20,technology21,technology21s,technology22,technology23,transport0,transport1,transport2,transport3,transport4,transport5,transport6,transport7,transport8,transport9,transport10,transport11,transport12,transport13}, a general transport formula that uniformly captures the effects of gain and loss has not yet been established. This limits our understanding of their impact on currents and hinders the prediction of novel nonequilibrium phenomena.

In this Letter, we derive a general formula to describe particle and energy currents in systems with arbitrary forms of gain or loss, providing a foundational framework for investigating how gain or loss impacts transport properties. Based on this formula, we reveal several nontrivial transport properties in open systems.

\begin{figure}[t]
	\centering
	\includegraphics[width=0.98\linewidth]{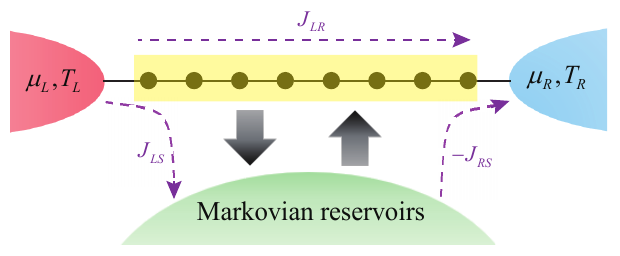}\\
	\caption{Model scheme: A central system (yellow region) is coupled to two leads (left and right) with temperatures $T_{L(R)}$ and chemical potentials $\mu_{L(R)}$. It also connects to Markovian reservoirs (green region) for particle transfer. The current splits into three components (Eq.~(\ref{current2})): $J_{LR}$, $J_{LS}$, and $J_{RS}$. $J_{LR}$ is described by the Landauer-B\"{u}ttiker formula, while the others arise from particle exchange with the reservoirs.}\label{01}
\end{figure}

{\em Model.---} We consider a one-dimensional system coupled to leads at its ends and exchanging particles with Markovian reservoirs (see Fig. \ref{01}). After tracing out the reservoirs, their effects are captured by gain/loss terms in the Lindblad master equation~\cite{UL3,GLindblad,HPBreuer,LMEQ1,LMEQ2,Brasil,SM} (see Supplemental Materials~\cite{SM} for derivation details)
\begin{equation}\label{1.1}
	\frac{d\rho}{dt}=-i[H,\rho]+\sum_{m}(2L_{m}\rho L_{m}^{\dagger}-\{L_{m}^{\dagger}L_{m},\rho\}),
\end{equation}
with the Hamiltonian 
\begin{equation}\label{Ham}
	H=H_{S}+\sum_{\alpha=L,R} (H_{\alpha}+H_{\alpha S}).
\end{equation} 
%Here, $H_S=\sum_{ij}h_{S,ij}c_{i}^{\dagger}c_{j}$ represents the Hamiltonian of the central system, where $c_j$ is the annihilation operator at site $j$.
Here, $H_S=\sum_{ij}(\mathbf{h_{S}})_{ij}c_{i}^{\dagger}c_{j}$ is the Hamiltonian of the central system, where $c_j$ is the annihilation operator at site $j$ and $(\mathbf{h_{S}})_{ij}$ is the $i$-th row and $j$-th column element of the matrix
$\mathbf{h_{S}}$, representing the hopping amplitude between sites $i$ and $j$.
The subscripts $\alpha=L$ and $\alpha=R$ denote the left and right leads, respectively, with the corresponding Hamiltonian $H_{\alpha}=\sum_k\epsilon_{\alpha,k}d_{\alpha,k}^{\dagger}d_{\alpha,k}$, 
where $\epsilon_{\alpha,k}$ is the energy of the 
$k$-th mode of the $\alpha$-lead and $d_{\alpha,k}^{\dagger}(d_{\alpha,k})$
creates (annihilates) a particle in this mode. The term $H_{\alpha S}=\sum_{jk}t_{\alpha,kj}d_{\alpha,k}^{\dagger}c_j+h.c.$ describes the coupling between the $\alpha$-lead and the system, where $t_{\alpha,kj}$ is the tunneling strength. The Lindblad operators $L_m$ in Eq.~(\ref{1.1}) describe particle gain or loss in the central system. Their form and strength depend on the system-reservoir coupling and reservoir correlation functions~\cite{SM}, and can generally be written as
\begin{equation} \label{gainloss}  
	L_{1,i}=\sum_{j}u_{ij}c_{j} \quad\textnormal{and} \quad L_{2,i}=\sum_{j}v_{ij}c_{j}^{\dagger}, 
\end{equation}
where $L_{1,i}$ and $L_{2,i}$ represent the loss and gain channels, respectively, which can be either on-site or non-local. 

{\em Generic formula.---} We now derive the particle and energy currents for the above system, denoted as 
$J^{0}$ and $J^{1}$, respectively. The particle (energy) currents flowing out of the left lead and into the right lead are, respectively, 
$J_{L}^{0}=-\frac{d\langle\mathcal{N}_{L}\rangle}{dt}$ ($J_{L}^{1}=-\frac{d\langle H_{L}\rangle}{dt}$) and $J_{R}^{0}=\frac{d\langle\mathcal{N}_{R}\rangle}{dt}$ ($J_{R}^{1}=\frac{d\langle H_{R}\rangle}{dt}$) ~\cite{SM}, where $\mathcal{N}_{\alpha}=\sum_kd^{\dagger}_{\alpha,k}d_{\alpha,k}$ and $\langle H_{\alpha}\rangle$ are the particle number and energy of the $\alpha$-lead, respectively. 
Using $J_L^{0}=-\frac{i}{\hbar}\langle[H,\mathcal{N}_L]\rangle$ and $J_L^{1}=-\frac{i}{\hbar}\langle[H,H_L]\rangle$, we can compute the currents flowing out of the left lead: 
\begin{equation}\label{current} 
	\begin{split}
		J_L^{0}=&\frac{1}{\hbar} \sum_{jk}(t_{L,kj}G_{SL,jk}^{<}-t_{L,kj}^{*}G_{LS,kj}^{<}), \\
		J_L^{1}=&\frac{1}{\hbar} \sum_{jk}\epsilon_{L,k}(t_{L,kj}G_{SL,jk}^{<}-t_{L,kj}^{*}G_{LS,kj}^{<}).
	\end{split}
\end{equation}

To trace out the leads, we use the Lindblad-Keldysh formalism~\cite{Keldysh1,Keldysh2,Keldysh3,Keldysh4}. The system's partition function can be written as $Z=\mathrm{tr}\rho(t)=\int D[\bar{\psi},\psi]e^{i\mathcal{S}[\bar{\psi},\psi]}$~\cite{SM}, where $\mathcal{S}$ is the Keldysh action and $\psi=(\psi^{+},\psi^{-})$ are Grassmann variables defined on the upper and lower branches of the Keldysh contour. This description contains redundancy~\cite{Keldysh1}, and for computational convenience, a Keldysh rotation is often performed: $\psi^{1}=\frac{1}{\sqrt{2}}(\psi^{+}+\psi^{-})$, $\psi^{2}=\frac{1}{\sqrt{2}}(\psi^{+}-\psi^{-})$, $\bar{\psi}^{1}=\frac{1}{\sqrt{2}}(\bar{\psi}^{+}-\bar{\psi}^{-})$, and
$\bar{\psi}^{2}=\frac{1}{\sqrt{2}}(\bar{\psi}^{+}+\bar{\psi}^{-})$. In this basis, the
Keldysh action $\mathcal{S}$ of the central region can be rewritten as~\cite{SM}
\begin{equation}\label{rho}
	\mathcal{S}=\int dt\left(
	\begin{array}{cc}
		\bar{\bm{\psi}}^{1} & \bar{\bm{\psi}}^{2} \\
	\end{array}
	\right)\left(
	\begin{array}{cc}
		i\partial_t-\mathbf{X} & i\mathbf{Y} \\
		0 & i\partial_t-\mathbf{X}^{\dagger} \\
	\end{array}
	\right)\left(
	\begin{array}{c}
		\bm{\psi}^{1} \\
		\bm{\psi}^{2} \\
	\end{array}
	\right),
\end{equation}
where $\mathbf{X}=\mathbf{h_S}-i\mathbf{P}-i\mathbf{Q}$ describes the damping dynamics~\cite{Prosen,Bardyn}, and $\mathbf{Y}=2(\mathbf{P}-\mathbf{Q})$ describes the imbalance between loss and gain. Note that the quantities in boldface are matrices. The elements of the matrices $\mathbf{P}$ and $\mathbf{Q}$ are $P_{jk}=\sum_{m}u_{mj}^{*}u_{mk}$ and $Q_{jk}=\sum_{m}v_{mj}v_{mk}^{*}$, corresponding to the loss and gain terms, respectively. The system's retarded, advanced and Keldysh Green's functions are obtained by inverting the matrix (\ref{rho}): $\mathbf{g}_S^{\mathcal{R}}=\frac{1}{\omega-\mathbf{X}}$,  $\mathbf{g}_S^{\mathcal{A}}=\frac{1}{\omega-\mathbf{X}^{\dagger}}$,  $\mathbf{g}_S^{\mathcal{K}}=-\frac{1}{\omega-\mathbf{X}}i\mathbf{Y}\frac{1}{\omega-\mathbf{X}^{\dagger}}$. Further combining the Langreth theorem~\cite{Keldysh4,Langreth}, after introducing the leads, one can obtain~\cite{SM}:
\begin{equation}\label{LTh2}
	\begin{split}
		\mathbf{G}_{S}^{>}=&i\mathbf{G}_{S}^{\mathcal{R}}(f_L\bm{\Gamma}_L+f_R\bm{\Gamma}_R-\bm{\Gamma}_L-\bm{\Gamma}_R-2\mathbf{P})\mathbf{G}_{S}^{\mathcal{A}}, \\ \mathbf{G}_{S}^{<}=& i\mathbf{G}_{S}^{\mathcal{R}}(f_L\bm{\Gamma}_L+f_R\bm{\Gamma}_R+2\mathbf{Q})\mathbf{G}_{S}^{\mathcal{A}},
	\end{split}
\end{equation}
where $f_{\alpha}=\frac{1}{e^{(\omega-\mu_{\alpha})/k_BT_{\alpha}}+1}$ is the Fermi distribution associated to the $\alpha$-lead, $\mathbf{G}_S^{\mathcal{R}}=(\mathbf{G}_S^{\mathcal{A}})^{\dagger}=[(\mathbf{g}_S^{\mathcal{R}})^{-1}-\tilde{\bm{\Sigma}}_{L}^{\mathcal{R}}-\tilde{\bm{\Sigma}}_{R}^{\mathcal{R}}]^{-1}$, and 
$\bm{\Gamma}_{\alpha}=i(\tilde{\bm{\Sigma}}_{\alpha}^{\mathcal{R}}-\tilde{\bm{\Sigma}}_{\alpha}^{\mathcal{A}})$ is the spectral density, with the lead self-energy being given by $\tilde{\bm{\Sigma}}_{\alpha}^{\mathcal{R}}=\bm{\Sigma}_{S\alpha}^{\mathcal{R}}\mathbf{g}_{\alpha}^{\mathcal{R}}\bm{\Sigma}_{\alpha S}^{\mathcal{R}}$ ($\alpha\in\{L,R\}$). 
The Langreth theorem also provides the relationship between the full Green's function ($\mathbf{G}$) and the bare Green's function ($\mathbf{g}$):
%By using the identities $\mathbf{g}_S^{>}=\frac{1}{2}(\mathbf{g}_S^{\mathcal{K}}+\mathbf{g}_S^{\mathcal{R}}-\mathbf{g}_S^{\mathcal{A}})$ and $\mathbf{g}_S^{<}=\frac{1}{2}(\mathbf{g}_S^{\mathcal{K}}-\mathbf{g}_S^{\mathcal{R}}+\mathbf{g}_S^{\mathcal{A}})$, the steady-state greater and lesser Green's functions can be derived as:
%\begin{equation}\label{greenf}
%	\mathbf{g}_S^{>}=-2i\mathbf{g}_S^{\mathcal{R}}\mathbf{P}\mathbf{g}_S^{\mathcal{A}},~~~
%	\mathbf{g}_S^{<}=2i\mathbf{g}_S^{\mathcal{R}}\mathbf{Q}\mathbf{g}_S^{\mathcal{A}}.
%\end{equation}
%By applying the Langreth theorem~\cite{Keldysh4,Langreth}, we obtain~\cite{SM}
\begin{equation}\label{LTh1}
	\begin{split}
		\mathbf{G}_{LS}^{<} =& \mathbf{g}_{L}^{<}\bm{\Sigma}_{LS}^{\mathcal{A}}\mathbf{G}_{S}^{\mathcal{A}}+\mathbf{g}_{L}^{\mathcal{R}}\bm{\Sigma}_{LS}^{\mathcal{R}}\mathbf{G}_{S}^{<} +\mathbf{g}_{L}^{\mathcal{R}}\bm{\Sigma}_{L}^{<}\mathbf{G}_{LS}^{\mathcal{A}}, \\
		\mathbf{G}_{SL}^{<} =& \mathbf{G}_{S}^{<}\bm{\Sigma}_{SL}^{\mathcal{A}}\mathbf{g}_{L}^{\mathcal{A}}+\mathbf{G}_{S}^{\mathcal{R}}\bm{\Sigma}_{SL}^{\mathcal{R}}\mathbf{g}_{L}^{<} +\mathbf{G}_{SL}^{\mathcal{R}}\bm{\Sigma}_{L}^{<}\mathbf{g}_{L}^{\mathcal{A}}.
	\end{split}
\end{equation}
By substituting Eq.~(\ref{LTh1}) into Eq.~(\ref{current}), the Green's functions of the system-lead coupling, $\mathbf{G}_{LS}$ and $\mathbf{G}_{SL}$, can be replaced by the system Green's function, $\mathbf{G}_{S}$. Further applying Eq.~(\ref{LTh2}), the current flowing out of the left lead, $J^{\lambda}_L$, can be derived~(see Supplemental Material~\cite{SM} for details), where $\lambda=0$ and $\lambda=1$ correspond to particle and energy currents, respectively.  A similar procedure gives the current flowing into the right lead,  $J^{\lambda}_R$. Due to the exchange of particles and energy between the central system and the environment, the currents flowing out of the left lead and into the right lead may differ, i.e.,
$J_{L}^{\lambda}\ne J_{R}^{\lambda}$ in the steady state.
The total current through the system is given by $J^{\lambda}=\frac{1}{2}(J_L^{\lambda}+J_R^{\lambda})$. 
These currents can be divided into three parts (see Fig. \ref{01}):
\begin{align} 
	&\quad \quad J^{\lambda}=J_{LR}^{\lambda}+J_{LS}^{\lambda}-J_{RS}^{\lambda},\label{current2} \\
	J_{LR}^{\lambda} =& \int\frac{d\omega}{2h}(f_L-f_R)\omega^{\lambda}\mathrm{Tr}[\bm{\Gamma}_{L}\mathbf{G}_{S}^{\mathcal{R}}\bm{\Gamma}_{R}\mathbf{G}_{S}^{\mathcal{A}} +\bm{\Gamma}_{R}\mathbf{G}_{S}^{\mathcal{R}}\bm{\Gamma}_{L}\mathbf{G}_{S}^{\mathcal{A}}], \nonumber \\
	J_{L(R)S}^{\lambda} &= \int\frac{d\omega}{h}f_{L(R)}\omega^{\lambda}\mathrm{Tr}[\bm{\Gamma}_{L(R)}\mathbf{G}_{S}^{\mathcal{R}}\mathbf{P}\mathbf{G}_{S}^{\mathcal{A}}] \nonumber\\ &+\int\frac{d\omega}{h}(f_{L(R)}-1)\omega^{\lambda}\mathrm{Tr}[\bm{\Gamma}_{L(R)}\mathbf{G}_{S}^{\mathcal{R}}\mathbf{Q}\mathbf{G}_{S}^{\mathcal{A}}]. \nonumber
\end{align}
The terms $J_{LR}^{\lambda}$, representing the Landauer-B\"{u}ttiker formula, and $J_{LS}^{\lambda}-J_{RS}^{\lambda}$
describe the current between two leads through the central system and through the reservoirs, respectively.
Here, $J_{L(R)S}$ consists of two parts: one describes the current from the leads to the loss channel through the central system, proportional to the particle distribution $f_{L(R)}$, and the other describes the current from the gain channel to the leads through the central system, proportional to the hole distribution $1-f_{L(R)}$.

{\em Applications.---} We will now discuss the interesting physical phenomena presented in Eq. (\ref{current2}). For convenience, the models we will discuss below as examples include only the nearest-neighbor hopping, i.e.,  $(\mathbf{h_{S}})_{ij}=-t_S \delta_{i,j\pm1}$, where $t_S$ is the hopping strength, and unless otherwise specified, we set
$t_S=1$.

\begin{figure}[t]
	\centering
	\includegraphics[width=1\linewidth]{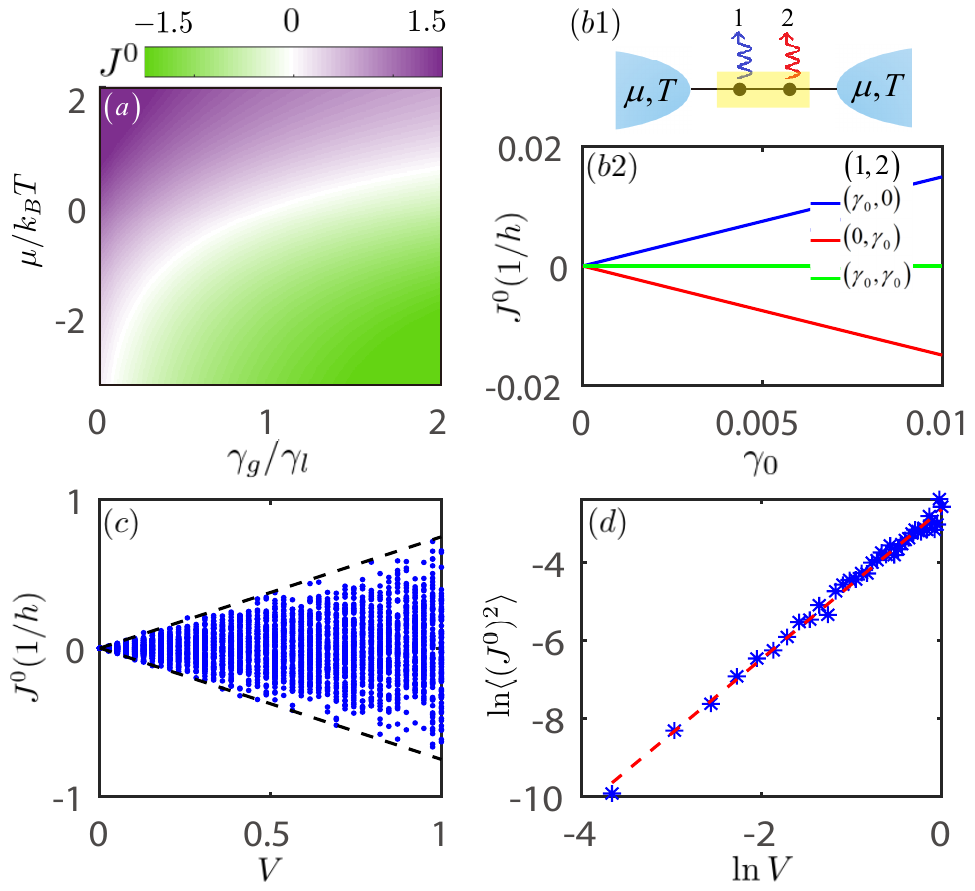}\\
	\caption{(a) $J^0$ as a function of $\gamma_{g}/\gamma_{l}$ and $\mu/k_BT$, in units of $1/h$, with $k_BT=20$, $\gamma_l=0.2$, and the system size $N=40$. (b1) Scheme of a two-site system with local monitoring. (b2) $J^0$ as a function of monitoring strength $\gamma_0$. The blue and red curves represent monitoring at the first and second sites, respectively, while the green curve represents monitoring at both sites. (c) $J^0$ as a function of disorder strength $V$ with $200$ samples and $N=100$. The black dashed line represents the envelope curve of the current. (d) The sample average of $(J^0)^2$. The red dashed line is the fitting function $\langle(J^0)^2\rangle\sim V^{1.91}$. Other parameters are (b-d) $\mu=0.1$, $k_BT=10^{-4}$, $\gamma_{l}=0.1$, and (a-d) $t_S=1$, $(\bm{\Gamma}_L)_{11}=(\bm{\Gamma}_R)_{NN}=1.1$.}\label{Fig2}
\end{figure}

To directly observe the effects of gain and loss on the current from Eq. (\ref{current2}), we assume 
$f_L=f_R=f$, meaning no chemical potential ($\mu_L=\mu_R=\mu$) or temperature ($T_L=T_R=T$) difference between the two leads, which results in no Landauer current $J^{\lambda}_{LR}$.  From Eq. (\ref{current2}), we see that  gain and loss can generate a current, except in two special cases, where the current vanishes. The first case occurs when the particle-to-hole ratio in the leads, $n_p/n_h=f/(1-f)$, matches the gain-to-loss ratio in the system, $\mathbf{Q}/\mathbf{P}$, such that the particle flow from the leads is exactly balanced by the gain and loss processes, leading to $J^{\lambda}_{LS}=J^{\lambda}_{RS}=0$ according to Eq. (\ref{current2}). In the high-temperature limit $k_BT\gg t_S$, the vanishing current condition simplifies to $\ln\frac{\mathbf{Q}}{\mathbf{P}}=\ln\frac{f}{1-f}=\frac{\mu}{k_BT}$.
Fig. \ref{Fig2} (a) illustrates how the particle current $J^0$ changes with $\mu/k_BT$ and $\gamma_g/\gamma_l$, where the gain and loss terms are assumed to be onsite and uniform, namely $L_{1,i}=\sqrt{\gamma_l}c_i$ and $L_{2,i}=\sqrt{\gamma_g}c_i^{\dagger}$. We observe that along the curve 
$\ln(\gamma_g/\gamma_l)=\mu/k_BT$, $J^0$ is zero, while on either side of this curve, $J^0$ is nonzero and opposite in direction.

The second case, in which the added gain and loss terms do not generate a current, is when the system and the gain and loss terms have inversion symmetry (IS), i.e., $\mathcal{P}^{-1}O_{ij}\mathcal{P}=O_{N-i+1,N-j+1}$, where the operator $\mathbf{O}$ represents the Hamiltonian $\mathbf{H}$, the loss matrix $\mathbf{P}$, and the gain matrix $\mathbf{Q}$~\cite{explain2}, and additionally, the spectral density matrix should satisfy $\mathcal{P}^{-1}\bm{\Gamma_{L}}\mathcal{P}=\bm{\Gamma_{R}}$.  It is easy to show from Eq. (\ref{current2}) that in this case,  $J^{\lambda}_{LS}=J^{\lambda}_{RS}$.  In Fig. \ref{Fig2} (b1), we show an example of two sites with a local monitoring described by $L_{1,j}=\sqrt{\gamma_j}c_{j}$. In the limit $\gamma_j\ll\omega$, the current can be written analytically as~\cite{SM}
\begin{equation}\label{current3} 
	J^0\approx(\gamma_1-\gamma_2)\Gamma\int\frac{d\omega}{h}f(\omega)\frac{|\omega+\frac{i\Gamma}{2}|^2-1 }{|(\omega+\frac{i\Gamma}{2})^2-1|^2},
\end{equation}
where $\Gamma=(\bm{\Gamma}_L)_{11}=(\bm{\Gamma}_R)_{NN}$.  When 
$\gamma_1=\gamma_2=\gamma_0$, i.e., both the system and the loss term preserve IS, the current 
$J^0$ remains zero regardless of $\gamma_0$, as shown by the green line in Fig. \ref{Fig2} (b2). Breaking IS by removing the loss on one site ($\gamma_1\neq\gamma_2$) induces a current proportional to $\gamma_1-\gamma_2$, as indicated by the red and blue lines in Fig. \ref{Fig2}(b2). Thus, when an inversion-symmetric system couples to an environment, even weak particle gain or loss can generate a current if it breaks IS. This property can be used to measure the characteristics of the environment. 
Lindblad loss and gain operators serve simultaneously as dissipation mechanisms and monitoring channels~\cite{monitor1,monitor2,monitor3}, and thus monitoring particles leaving or entering the system also affects the system's dynamics. Our results may provide valuable insights into transport phenomena across various monitored quantum platforms.

On the other hand, if the gain and loss terms preserve IS but the system does not, a current can also arise. We consider uniform onsite loss to preserve IS in the dissipation term, and introduce a disorder potential $\sum_{j}V_jc_{j}^{\dagger}c_{j}$ to break IS in the system Hamiltonian $H_S$, with $V_j$ randomly distributed in $[-V, V]$. In one dimension, all eigenstates are localized even for arbitrarily weak disorder. As shown in Fig. \ref{Fig2}(c), the current $J^0$ remains zero at $V=0$ due to IS, but grows with increasing $V$, exhibiting strong sample-to-sample fluctuations. These behaviors can be analytically understood using a two-site model (see Supplemental Materials~\cite{SM}): in the weak disorder limit, $J^0\approx \frac{c}{2} (V_1-V_2)$~\cite{SM}, where $c$ depends on the loss strength $\gamma$ and the Fermi distribution. $J^0$ thus vanishes for symmetric potentials $V_1=V_2$, and reaches $\pm cV$ for maximal asymmetry  $V_1=-V_2=\pm V$, forming linear envelope bounds. In the strong disorder regime, resonant tunneling dominates, and the envelope of $J^0$ fluctuations saturates to a $V$-independent value~\cite{SM}. This crossover reflects  competition between dissipation and disorder. Fig. \ref{Fig2}(d) further shows that the sample average of  $(J^0)^2$ scales with disorder strength as $\langle(J^0)^2\rangle\sim V^2$. 
In contrast to closed systems, where disorder suppresses transport, here it can induce current. With spin-dependent disorder, one can further generate spin-polarized currents or even pure spin currents without net charge flow~\cite{SM}.

\begin{figure}[t]
	\centering
	\includegraphics[width=1\linewidth]{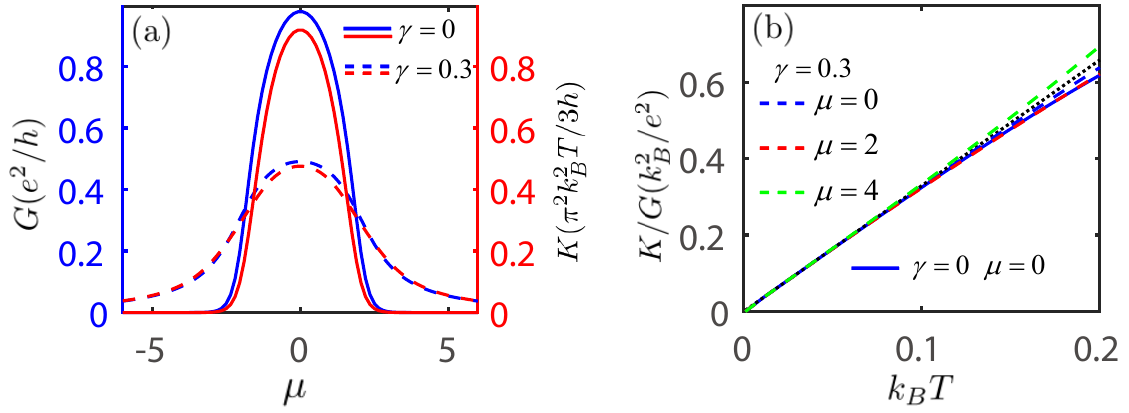}\\
	\caption{(a) $G$ and $K$ as functions of the chemical potential $\mu$ at temperature $k_B T = 0.2$, where $K$ is expressed in units of the thermal conductance quantum $\frac{\pi^2 k_B^2 T}{3h}$. (b) The ratio $K/G$ as a function of $T$. $\mu = 0$, $\mu = 2$, and $\mu = 4$ are located inside the band, at the band edge, and outside the band, respectively. The black dotted line represents the Wiedemann-Franz law. Here we take $t_S=1$ and $N = 80$.}\label{Fig3}
\end{figure}

We then consider the current when there is a chemical potential and temperature gradient between the leads, i.e., $f_L\neq f_R$.
We set $\mu_R=\mu$, $T_{R}=T$, $f_{R}=f$, and $\mu_{L}=\mu+\delta\mu$, $T_{L}=T+\delta T$, $f_{L}=f+\delta f$. For convenience, we rewrite the current in Eq. (\ref{current2}) as two parts:
\begin{equation}\label{currentX} 
	J^{\lambda}=J_0^{\lambda}+\delta J_1^{\lambda}, 
\end{equation}
where $J_0^{\lambda}$ denotes the net current under equilibrium conditions ($\delta f=0$), and $\delta J_1^{\lambda}$, referred to as the response current, describes the current induced when $\delta f\neq 0$. From Eq. (\ref{current2}), it is easy to obtain
\begin{equation}\label{5.1}
	\delta J_1^{\lambda}=\int\frac{d\omega}{h}\omega^{\lambda}\delta f(\omega)\tau_1(\omega),
\end{equation}
where $\tau_1 = \tau_{11} + \tau_{12}$, with  $\tau_{11} = \frac{1}{2}\mathrm{Tr}[\bm{\Gamma}_{L}\mathbf{G}_{S}^{\mathcal{R}}\bm{\Gamma}_{R}\mathbf{G}_{S}^{\mathcal{A}} + \bm{\Gamma}_{R}\mathbf{G}_{S}^{\mathcal{R}}\bm{\Gamma}_{L}\mathbf{G}_{S}^{\mathcal{A}}]$ describing direct transmission between the leads and $\tau_{12} = \mathrm{Tr}[\bm{\Gamma}_{L}\mathbf{G}_{S}^{\mathcal{R}}(\mathbf{P}+\mathbf{Q})\mathbf{G}_{S}^{\mathcal{A}}]$ accounting for indirect transmission via the reservoirs. For inversion-symmetric systems, $J_0=0$ and the current $J=\delta J_1$ follows thermodynamic gradients (from high to low $\mu$ and $T$). When IS is broken, $J_0$ and $\delta J_1$ compete, adding constructively when aligned or being dominated by the larger term when opposed. 

The electrical conductance $G$, determined by the response current $\delta J^{0}_1$, depends on the transmission function $\tau_1$ via $G=\frac{e^2}{h}\int d\omega\frac{\delta f(\omega)}{\delta\mu}\tau_1(\omega)$~\cite{SM}. The size dependence of $G$ is encoded in $\tau_1$: without gain and  loss ($\tau_{12}=0$), transport is ballistic and $\tau_{11}$ is size-independent; with gain or loss, $\tau_{11}$ decays exponentially due to particles scattering into reservoirs, while $\tau_{12}$, which arises from reservoir channels whose number increases with system size, remains nearly  size-independent (see End Matter). Thus, in the thermodynamic limit, the conductance is dominated by indirect transmission~\cite{explainF}.

The transmission function $\tau_1$ also determines the thermal conductance $K$~\cite{Loc11,SM,Butcher}. In the low-temperature limit, the ratio $K/G$ follows the Wiedemann-Franz law $K/G\approx\mathcal{L}T$, with $\mathcal{L}=\frac{\pi^2}{3}(\frac{k_B}{e})^2$ being the Lorenz number. Without dissipation, both $G$ and $K$ are finite only within the energy band [Fig. \ref{Fig3}(a)], and the law holds only there [Fig. \ref{Fig3}(b)]~\cite{explainKG}. We then consider the case where each site is subject to equal loss and gain ($\gamma_l=\gamma_g=\gamma$), which causes the conductances to broaden and smooth out near the band edge [Fig. \ref{Fig3}(a)], allowing the Wiedemann-Franz law to hold both inside and outside the band [Fig. \ref{Fig3}(b)].

\begin{figure}[t]
	\centering
	\includegraphics[width=1\linewidth]{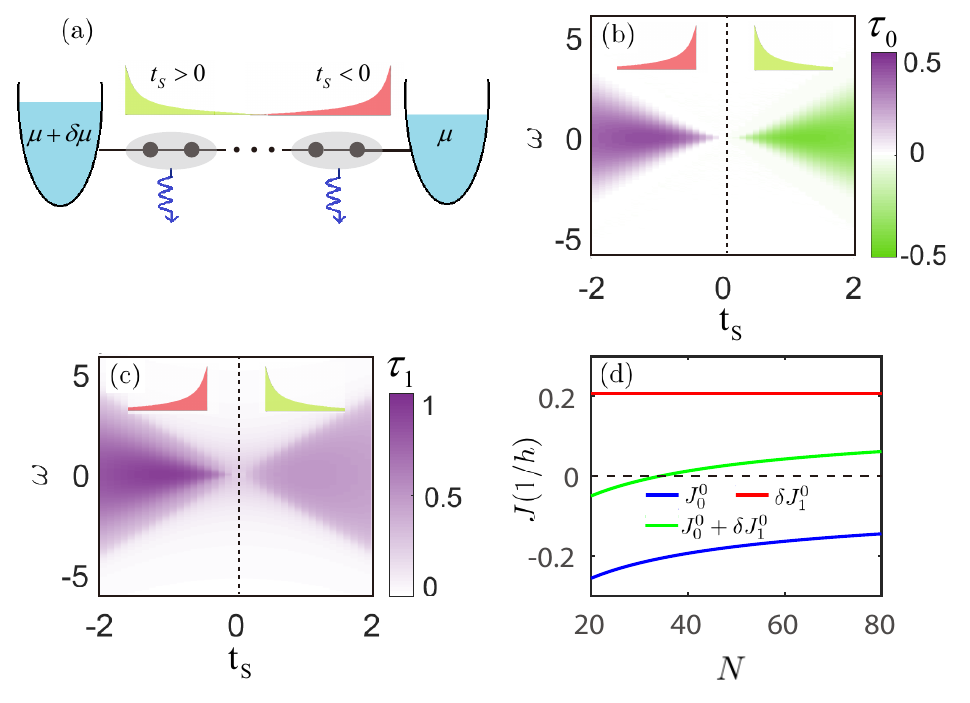}\\
	\caption{(a) Schematic of the setup to study transport in a system with NHSE caused by bond loss. The transmission functions (b) $\tau_0$ and (c) $\tau_1$ as a function of $t_S$ and $\omega$ for a system with size $N = 80$.
		(d) The variation of $J_0^0$, $\delta J_1^0$, and $J^0 = J_0^0 + \delta J_1^0$ with system size, with $t_S=3$ fixed. Other parameters: (b-d) $\gamma_{-}=1$, $\mu=0.12$; (c-d) $\delta\mu=0.6$.}\label{Fig4}
\end{figure}

Eq. (\ref{current2}) also applies to non-onsite gain or loss.
As an example, we consider the bond-loss $L_{j}=\sqrt{\gamma_{-}}(c_j-ic_{j+1})$, which induces the non-Hermitian skin effect (NHSE) in this system: all  eigenstates localize at the left (right) end for $t_S>0$ ($t_S<0$)~\cite{NHSE1,NHSE2} [Fig. \ref{Fig4}(a)].
In the zero-temperature limit, the current is  $J_0^0=\int_{-\infty}^{\mu} \frac{d\omega}{h} \tau_0(\omega)$ and $\delta J_1^{0} = \int_{\mu}^{\mu+\delta\mu}\frac{d\omega}{h}\tau_1(\omega)$, where $\tau_0$ and $\tau_1$ are the transmission functions~\cite{SM}. Figures \ref{Fig4}(b) and (c) show $\tau_0$ and $\tau_1$ as functions of frequency $\omega$ and $t_S$, respectively. It can be seen that $\tau_0$ is positive (negative) when skin modes localize on the right (left) end, so $J_0^0$ flows in the same direction as the skin modes when $\mu_L = \mu_R$. Under a voltage bias  ($\delta \mu \neq 0$), the response current $\delta J_1^0$  always flows from high to low $\mu$, as shown in Fig. \ref{Fig4}(c). Thus, if the bias opposes the skin direction, the direction of the total current depends on the competition between $J_0^0$ and $\delta J_1^0$. In the thermodynamic limit, $J_0^0$ decreases with increasing system size and, for $|t_S|\geq|\gamma_{-}|$, satisfies: $J_0^0 \sim N^{-0.5}$ for $\mu \geq 0$ and $J_0^0 \sim \frac{e^{-N}}{N}$ for $\mu < 0$~\cite{SM}. This implies that the current corresponding to the NHSE vanishes as the system size tends to infinity, while $\delta J_1^0$ does not depend on the system size [Fig. \ref{Fig4}(d)]. Therefore, the direction of the current $J^0$ can be changed by altering the system size to change $J_0^0$ or by changing $\delta \mu$ to adjust $\delta J_1^0$.

%The whole system will evolve to the universal identity density matrix $\hat{\rho}_{ss}=\hat{I}/L$ with the lattice length $L$, if all the jump operators $\hat{O}_{j}$ are Hermitian, namely, when $\alpha=\pi/2$ \cite{xx}. Those considering dissipators in Eq. (\ref{jumpoperator}) breaks the relevant symmetries of system and makes the steady state $\hat{\rho}_{ss}$ uniqueness \cite{xx} [explain this in ref. or SM]. 

{\em Discussion and conclusion.---} We have derived a general formula for the current through a region with particle exchange with the environment, and further demonstrate its usefulness by identifying several novel transport features: (I) A current is generated when gain or loss terms lack inversion symmetry; (II) Disorder can induce current; (III) Gain and loss make the thermal and electrical conductances continuous and cause their ratio to satisfy the Wiedemann-Franz law inside and outside the energy band; (IV) The bond loss-induced NHSE and the current caused by the chemical potential difference exhibit an interesting competition. Our work establishes a versatile framework for studying the influence of particle gain or loss on transport phenomena, opening avenues for future studies of open and non-Hermitian physics.

\begin{acknowledgments}
	This work is supported by National Key R\&D Program of China under Grant No.2022YFA1405800, the Key-Area Research and Development Program of Guangdong Province (Grant No.2018B030326001), Guangdong Provincial Key Laboratory(Grant No.2019B121203002).
\end{acknowledgments}

%%%%%%%%%%%%%%%%%%%%%%%%%%%%%%%%%%%%%%%%%%%%%

%%%%%%%%%%%%%%%%%%%%%%%%%%%%%%%%%%%%%%%%%%%%%%%%%%%%%
\section{End Matter}
%This End Matter examines two key aspects under the influence of gain and loss: the scaling behavior of conductance with system size, and the formulation of transport for bosonic systems.

\begin{figure}[t]
	\centering
	\includegraphics[width=1\linewidth]{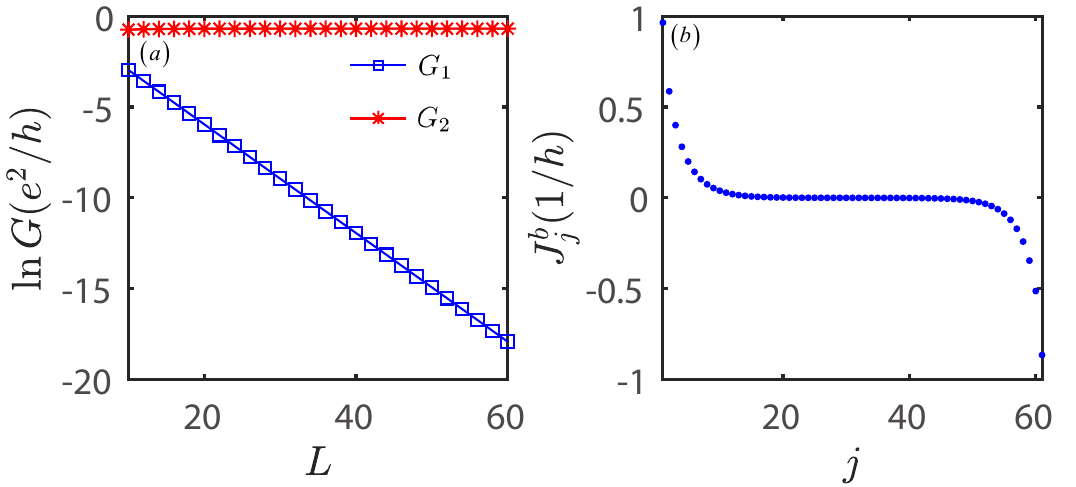}\\
	\caption{(a) Zero-temperature conductance as a function of system size.
		(b) Spatial profile of the bond current $J^b_{j}$ at zero temperature.
		Parameters: $t_S=1$, $\gamma_l=0.2$, $\gamma_g=0.1$, $\mu_L=0.2$, and $\mu_R=0.1$.}\label{FigS3} 
\end{figure}

{\em Conductance Scaling with System Size in the Presence of Gain and Loss.---} We now consider the size dependence of conductance. For convenience, we focus on the zero-temperature limit, such that the conductance is proportional to the transmission function, $G = \frac{e^2}{h}\tau_1$.
As discussed above, we decompose the transmission function into two contributions, $\tau_1 = \tau_{11} + \tau_{12}$,
leading to $G_1 = \frac{e^2}{h}\tau_{11}$ and $G_2 = \frac{e^2}{h}\tau_{12}$, respectively.
Since $\tau_{12} = \text{Tr}\!\left[\bm{\Gamma}_{L}\mathbf{G}_{S}^{\mathcal{R}}(\mathbf{P}+\mathbf{Q})\mathbf{G}_{S}^{\mathcal{A}}\right]$,
the contribution $G_2$ arises solely from the presence of loss ($\mathbf{P}$) and gain ($\mathbf{Q}$). In their absence, $G_2$ vanishes. Moreover, $\tau_{11}$ involves the retarded Green's function $\mathbf{G}_S^{\mathcal{R}}$, whose inverse reads
$(\mathbf{G}_S^{\mathcal{R}})^{-1} = \omega - \mathbf{h}_S + i\mathbf{P} + i\mathbf{Q} - \tilde{\bm{\Sigma}}_{L}^{\mathcal{R}} - \tilde{\bm{\Sigma}}_{R}^{\mathcal{R}}$.
Therefore, introducing loss or gain into the system not only generates the conductance contribution $G_2$, but also modifies $G_1$, owing to the effective change in the system Hamiltonian from $\mathbf{h}_S$ to $\mathbf{h}_S - i\mathbf{P} - i\mathbf{Q}$.
To illustrate, consider the simplest tight-binding model, $H_S=-t_S\sum_{j}(c_j^{\dagger}c_{j+1}+h.c.)$. In the absence of loss and gain, the finite conductance comes from the contacting conductance at the boundary, and the bulk is resistenceless. 
Without loss of generality, we consider uniform particle loss and gain, with strengths $\gamma_l$ and $\gamma_g$, applied at each site, i.e., $L_{1,j} = \sqrt{\gamma_l}c_j$ and $L_{2,j} = \sqrt{\gamma_g}c_j^{\dagger}$. As shown in Fig. \ref{FigS3}(a), $G_1$ decays exponentially with system size, whereas $G_2$ remains constant. Due to the presence of gain and loss at each lattice site, particles acquire a finite lifetime in the bulk, which results in a finite penetration depth (determined by the strength of gain and loss). As illustrated in Fig. \ref{FigS3}(b), the spatial profile of the bond current exhibits exponential decay from the edges toward the center. Here, the bond current $J_{j}^{\mathrm{b}}$ refers to the particle flow from site $j-1$ to site $j$, and is defined through the coherent hopping: $\frac{i}{\hbar}\langle[\mathcal{H},c_j^{\dagger}c_j]\rangle=J_{j}^{\mathrm{b}}-J_{j+1}^{\mathrm{b}}$, with the boundary conditions $J_{1}^{\mathrm{b}} = J^0_L$ and $J_{L+1}^{\mathrm{b}} = J^0_R$. 

The contribution $G_1$ originates from the direct transmission term $\tau_{11}$, corresponding to the process in which particles propagate directly from one lead to the other. This process requires particles to traverse the entire system.  Particles in the bulk have a finite lifetime, leading to an exponential suppression of the transmission probability $\tau_{11}$ with system size $L$. Consequently, $G_1$ decreases exponentially as $L$ increases. By contrast, $G_2$ arises from the indirect transmission term $\tau_{12}$, in which particles injected from one lead undergo scattering via gain/loss channels coupled to the reservoirs within the bulk before entering the other lead. In this process, particles do not need to traverse the whole system. As a result, the contribution to $G_2$ mainly comes from regions near the two boundaries, within a distance of the penetration depth from the leads. Increasing the system size (as long as $L$ exceeds the penetration depth) neither enlarges nor reduces the effective scattering region (or the number of available channels). Therefore, $G_2$ remains independent of system size. As a result, in the long-system limit, the total conductance approaches a constant value $G\approx G_2 \neq 0$, demonstrating that non-zero conductance persists even in the thermodynamic limit.

It is worth noting that the above argument concerning the size dependence of $G_1$ and $G_2$ holds in general, but exceptions do exist. For example, in the presence of the non-Hermitian skin effect, as demonstrated earlier, the transmission function from the left (right) lead to the right (left) lead, $\text{Tr}[\bm{\Gamma}_{L}\mathbf{G}_{S}^{\mathcal{R}}\bm{\Gamma}_{R}\mathbf{G}_{S}^{\mathcal{A}}] (\text{Tr}[\bm{\Gamma}_{R}\mathbf{G}_{S}^{\mathcal{R}}\bm{\Gamma}_{L}\mathbf{G}_{S}^{\mathcal{A}}])$, can remain finite and not decay with system size if the eigenmodes of the effective Hamiltonian are skin-localized near the right (left) boundary. Consequently, both $G_1$ and $G_2$ can contribute in the thermodynamic limit.

{\em Transport in Bosonic Systems with Gain and Loss.---} In previous studies of transport phenomena, the focus has primarily been on fermionic systems. However, with the advancement of experiments in cold-atom and photonic systems, bosonic systems have also become increasingly relevant. In the presence of gain and loss, and following a derivation analogous to that for fermionic systems (see the Supplementary Materials for detailed derivations), we obtain the transport formula for bosonic systems. The steady-state current can likewise be expressed in the form of Eq. (\ref{current2}), where the Landauer term takes exactly the same structure as in the fermionic case, namely, $J_{LR}^{\lambda} = \int\frac{d\omega}{2h}(\tilde{f}_L-\tilde{f}_R)\omega^{\lambda}\mathrm{Tr}[\bm{\Gamma}_{L}\mathbf{G}_{S}^{\mathcal{R}}\bm{\Gamma}_{R}\mathbf{G}_{S}^{\mathcal{A}} +\bm{\Gamma}_{R}\mathbf{G}_{S}^{\mathcal{R}}\bm{\Gamma}_{L}\mathbf{G}_{S}^{\mathcal{A}}]$, with the only difference being that $\tilde{f}_{\alpha} = \frac{1}{e^{(\omega - \mu_\alpha)/k_B T_\alpha} - 1}$ now denotes the bosonic distribution. By contrast, the contribution to the current associated with gain and loss takes a slightly modified form:
\begin{equation}\label{Boson}
	\begin{split}
		J_{L(R)S}^{\lambda} &= \int\frac{d\omega}{h}\tilde{f}_{L(R)}\omega^{\lambda}\mathrm{Tr}[\bm{\Gamma}_{L(R)}\mathbf{G}_{S}^{\mathcal{R}}\mathbf{P}\mathbf{G}_{S}^{\mathcal{A}}] \\ &-\int\frac{d\omega}{h}(\tilde{f}_{L(R)}+1)\omega^{\lambda}\mathrm{Tr}[\bm{\Gamma}_{L(R)}\mathbf{G}_{S}^{\mathcal{R}}\mathbf{Q}\mathbf{G}_{S}^{\mathcal{A}}]. 
	\end{split}
\end{equation}
We emphasize two important points: 1. The above transport formula is valid only when $\omega > \mu_{\alpha}$; otherwise, the system undergoes Bose-Einstein condensation. 2. The gain term ($\mathbf{Q}$) behaves differently for bosons and fermions. For fermions, Pauli blocking ensures that the current from the gain channel is proportional to the hole density $(1-f)$, i.e., it is suppressed by occupation. For bosons, in contrast, the current is proportional to $(\tilde{f}+1)$, implying that higher occupation in the lead facilitates further injection. This is the hallmark of bosonic stimulation (also known as Bose enhancement).

\begin{figure}[h]
	\centering
	\includegraphics[width=0.65\linewidth]{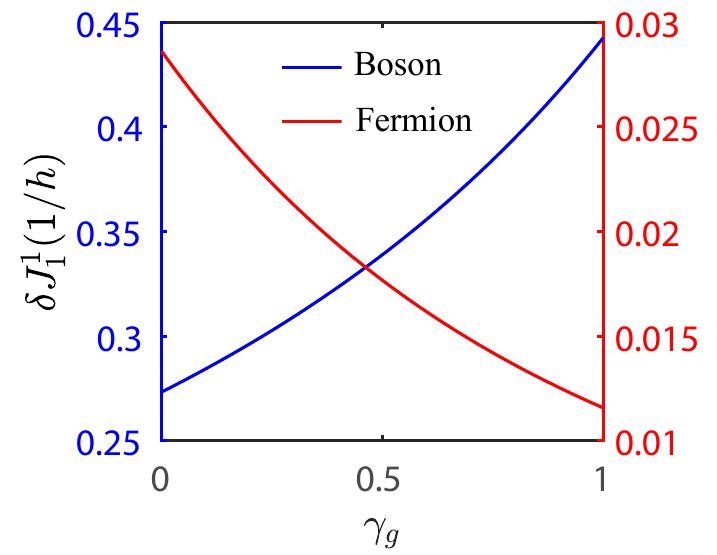}\\
	\caption{Response energy current $\delta J_{1}^{1}$ with $\epsilon_0 = 0.3$ as a function of the gain rate $\gamma_g$. 
		The blue curve corresponds to bosons and the red curve to fermions. 
		The system parameters are temperature $T = 0.2$, temperature bias $\delta T = 0.01$, and chemical potential $\mu_L=\mu_R = 0$. 
		The loss rate is fixed at $\gamma_l = 0.1$.}\label{BF}
\end{figure}

To gain direct insight into the difference in transport properties between bosonic and fermionic systems induced by gain, 
we consider a toy model consisting of a single mode, 
$H_S=\epsilon_0 c^{\dagger}c$, with a fixed loss rate $\gamma_l$ and a tunable gain rate $\gamma_g$. 
Similar to the treatment in fermionic systems, the total current is decomposed into two parts, 
$J=J_0+\delta J_1$, where $J_0$ denotes the equilibrium current and the response current is given by
$\delta J_1^{\lambda} = \int \frac{d\omega}{h}\,\omega^{\lambda}\delta\tilde{f}(\omega)\tau_1(\omega)$,
with the transmission function expressed as
$\tau_1(\omega) = \text{Tr}\Bigg[
\frac{1}{2}\bm{\Gamma}_{L}\mathbf{G}_{S}^{\mathcal{R}}\bm{\Gamma}_{R}\mathbf{G}_{S}^{\mathcal{A}}
+ \frac{1}{2}\bm{\Gamma}_{R}\mathbf{G}_{S}^{\mathcal{R}}\bm{\Gamma}_{L}\mathbf{G}_{S}^{\mathcal{A}}
+ \bm{\Gamma}_{L}\mathbf{G}_{S}^{\mathcal{R}}(\mathbf{P}-\mathbf{Q})\mathbf{G}_{S}^{\mathcal{A}}
\Bigg]$.
By applying a small temperature bias $\delta T$, we evaluate the current $\delta J_1^1$. 
As shown in Fig. \ref{BF}, $\delta J_1^1$ increases in the bosonic case with growing $\gamma_g$, 
since the enhanced particle number in the system promotes transport. 
In contrast, for fermions $\delta J_1^1$ decreases as the increasing occupation suppresses transport due to Pauli blocking.

{\em Relation to B\"{u}ttiker's Voltage-Probe Approach.---} The relationship between this work and B\"{u}ttiker's seminal paper~\cite{ButtikerVP} warrants clarification. B\"{u}ttiker's voltage-probe approach provides an elegant scattering-theory framework for incorporating decoherence into quantum transport. By attaching fictitious reservoirs (probes) to the conductor and tuning their chemical potentials to enforce a zero-net-current condition, it effectively simulates a phase-randomizing inelastic scatterer while strictly conserving particle number. In this sense, it offers a phenomenological route to introduce dephasing within the Landauer-B\"{u}ttiker formalism.

The voltage-probe model contains no explicit environmental degrees of freedom and, by construction, cannot describe particle-nonconserving processes. By contrast, we focus on dissipation arising from environment-induced particle gain and loss. Using a Lindblad-Keldysh framework with explicit system-environment coupling, we obtain genuine open-system dynamics that incorporate particle exchange, energy flow, and non-unitary evolution of the reduced density matrix. Consequently, environmental properties enter our transport formulas explicitly, enabling phenomena that lie beyond the scope of voltage-probe models, such as current generation in symmetric setups with asymmetric gain/loss, non-Hermitian skin-induced transport, and size-independent conductance contributions in the large-system limit.

In summary, the two approaches differ in motivation, methodology, and physical implications. B\"{u}ttiker's framework is well suited for describing dephasing-related phenomena arising from inelastic scattering, whereas the microscopic framework developed here provides a unified and systematic description of quantum transport in the presence of particle gain and loss. The two approaches should therefore be viewed as complementary, each tailored to distinct physical regimes.

%%%%%%%%%%%%%%%%%%%%%%%%%%%%%%%%%%%%%%%%%%%%%%%%%%%%%
%%%%%%%%%%%%%%%%%%%%%%%%%%%%%%%%%%%%%%%%%%%%%%%%%%%%%%%
\global\long\def\id{\mathbbm{1}}
\global\long\def\ui{\mathbbm{i}}
\global\long\def\ud{\mathrm{d}}
%%%%%%%%%%%%%%%%%%%%%%%%%%%%%%%%%%%%%%%%%%%%%%%%%%%%%%%%%%%%%%%
\setcounter{equation}{0} \setcounter{figure}{0}
\setcounter{table}{0} %\setcounter{page}{1} \makeatletter
\renewcommand{\theparagraph}{\bf}
\renewcommand{\thefigure}{S\arabic{figure}}
\renewcommand{\theequation}{S\arabic{equation}}

\onecolumngrid
\flushbottom
%%%%%%%%%%%%%%%%%%%%%%%%%%%%%%%%%%%%%%%%%%%%%%%%%
\newpage
\section*{\large Supplementary Material:\\Extended Landauer-B\"{u}ttiker Formula for Current through Open Quantum Systems with Gain or Loss}
%\section*{\normalsize SUPPLEMENTAL MATERIAL}

In the Supplementary Materials, we first clarify the distinction between Markovian reservoirs and conventional leads, and show the derivation of the Lindblad master equation. We then provide the definition of the current, introduce the Keldysh theory of open fermionic systems, and present the detailed derivation of the extended Landauer-B\"{u}ttiker formula in fermionic and bosonic systems. Next, we analytically analyze the effect of inversion symmetry in gain or loss terms on the current using a two-site model, and examine how disorder-induced symmetry breaking leads to current generation. Finally, we study the impact of gain or loss on the Wiedemann-Franz law and investigate the behavior of the current in the presence of the skin effect.

\section{I. Comparative analysis: Markovian reservoirs and conventional leads}
In this section, we clarify the key difference between the Markovian reservoirs responsible for gain and loss and the conventional leads. The system is coupled at both ends to two leads, which serve as particle reservoirs. The particle exchange between the system and the leads constitutes coherent transport, where quantum phase information is preserved. %Accordingly, the combined Hamiltonian of the system and leads can be written as Eq. (2) in our manuscript: $H=H_{S}+\sum_{\alpha=L,R} (H_{\alpha}+H_{\alpha S})$, where the term $H_{\alpha S}=\sum_{jk}t_{\alpha,kj}d_{\alpha,k}^{\dagger}c_j+h.c.$ describes the coherent coupling between the $\alpha$-lead and the central system. 
Since the process of particles entering the system from one lead, traversing the system, and exiting into the other lead can be fully described by coherent scattering theory (e.g., Landauer-B\"{u}ttiker formalism), there is no need to invoke the Lindblad master equation in this case.

In contrast, the gain or loss mechanisms represent incoherent particle exchange with effective Markovian reservoirs. In these processes, quantum coherence (i.e., phase information) is destroyed or no longer tracked. As an intuitive picture, the particles in the system initially possess well-defined coherent phase relations. When a particle escapes into a Markovian reservoir--with no record of its exit time or destination--the coherence among the remaining particles is disrupted. Conversely, a particle injected from the reservoir carries random phase information, further destroying the system's original phase coherence. This memoryless reservoir is not explicitly modeled as a third terminal with a well-defined chemical potential or Fermi distribution. Instead, particle injection and extraction are described by the Lindblad master equation using jump operators, as given in Eq. (3) in the main text: $L_{1,i}=\sum_{j}u_{ij}c_{j}$ and  $L_{2,i}=\sum_{j}v_{ij}c_{j}^{\dagger}$. 
These processes are incoherent and irreversible, with no information retained about the origin or destination of the exchanged particles. Only when the reservoir satisfies the Markovian condition can its effects be captured by a Lindblad master equation after tracing out its degrees of freedom (as can be seen in the later derivation).
Due to the loss of coherence, the transport behavior of the system cannot be described by coherent scattering theory.
Thus, the standard Landauer-B\"{u}ttiker formula, based on coherent scattering, necessitates modifications and extensions. 

\begin{figure}[htbp]
	\centering
	\includegraphics[width=0.4\linewidth]{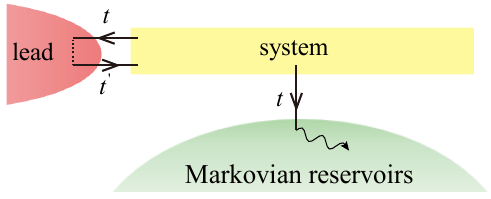}\\
	\caption{Scheme of particles scatter into the leads and Markovian reservoirs.}\label{FigSS1}
\end{figure}

The above discussion can be further interpreted as a second-order process, as illustrated in Fig. \ref{FigSS1}. A particle 
scatters from the system into the left lead at time $t$, propagates within the lead from $t$ to $t^{'}$, and then scatters back into the system. The process can be described by
\begin{equation}\label{R.1}
	G_S^{(2)}(t,t^{'})\sim g_S^{(0)}(t)\Sigma_{SL}g_{L}^{(0)}(t-t^{'})\Sigma_{LS}g_S^{(0)}(t^{'}),
\end{equation}
where $g^{(0)}$ denotes the bare Green's function without coupling to leads and reservoirs. We see that it contains the Green's function in the lead $g_{L}^{(0)}(t-t^{'})$, which can be time-dependent and hence carry the memory effect of leads. As a comparison, when a particle scatter into the Markovian reservoir, it will thermalize/dissipate rapidly and lose its initial state information.

\section{II. Derivation of the Lindblad master equation}
In this section, we present the derivation and examine the validity of the Lindblad master equation. For simplicity, we omit the degrees of freedom associated with the leads, as they do not influence the results. The total Hamiltonian of the system, including the Markovian reservoirs, is given by
\begin{equation}\label{S.1.1}
	H=H_{S}+\sum_{m}[H_{B}^{(m)}+H_{SB}^{(m)}],
\end{equation}
with $H_{S}$ represents the Hamiltonian of the central system, $H_{B}^{(m)}$ denotes the Hamiltonian of the $m$-th reservoir, and $H_{SB}^{(m)}$ describes the coupling between the central system and the $m$-th reservoir. The Hamiltonian of the central system, $H_{S}(c_j;c_j^{\dagger})$, is expressed as a function of the annihilation operators ($c_j$) and creation operators ($c_j^{\dagger}$) at site $j$. Each reservoir is modeled as Fermionic (qausi-)particles with the Hamiltonian $H_{B}^{(m)}=\sum_{k}\epsilon_k^{m}a_{k}^{(m)\dagger}a_{k}^{(m)}$, where $a_{k}^{(m)}(a_{k}^{(m)\dagger})$ denotes the annihilation(creation) operator of $m$-th reservoir and $\epsilon_k^{m}$ represents its energy. The coupling between central system and the $m$-th reservoir is given by
\begin{equation}\label{S.1.2}
	H_{SB}^{(m)}=\sum_{jk}[\lambda_{jk}^{(m)}a_{k}^{(m)\dagger}c_{j}+\lambda_{jk}^{(m)*}c_{j}^{\dagger}a_{k}^{(m)}].
\end{equation}
where $\lambda_{jk}^{(m)}$ denotes the coupling strength. %It is assumed that the reservoirs satisfy the Markovian condition, meaning the relaxation time $\tau_B$ of reservoirs is significantly shorter than the dynamical evolution timescale $\Delta t$ of the central systems \cite{Citro,Brasil}.

The evolution of density matrix follows the Liouville-von Neumann equation $\frac{d\rho}{dt}=-i[H,\rho]$. To facilitate calculations, we adopt the interaction picture. For any operator $O$ (such as $c_j(c_j^{\dagger})$, $a_k^{(m)}(a_k^{(m)\dagger})$) in the Schr\"{o}dinger picture, its interaction picture representation is defined as
$\bar{O}(t)=e^{i[H_S+\sum_mH_B^{(m)}]t}O e^{-i[H_S+\sum_mH_B^{(m)}]t}$. The density matrix in the interaction picture is defined as $\bar{\rho}(t)=e^{i[H_S+\sum_mH_B^{(m)}]t}\rho(t) e^{-i[H_S+\sum_mH_B^{(m)}]t}$. The coupling term in the interacting picture is expressed as
\begin{equation}\label{S.1.3}
	\bar{H}_{SB}^{(m)}(t)=\sum_{jk}[\lambda_{jk}^{(m)}\bar{a}_{k}^{(m)\dagger}(t)\bar{c}_{j}(t) +\lambda_{jk}^{(m)*}\bar{c}_{j}^{\dagger}(t)\bar{a}_{k}^{(m)}(t)].
\end{equation}
The Liouville equation in the interacting picture then becomes
\begin{equation}\label{S.1.4}
	\frac{d\bar{\rho}(t)}{dt}=-i[\sum_m\bar{H}_{SB}^{(m)}(t),\bar{\rho}(t)].
\end{equation}
Over a time interval $\Delta t$, the change in the density matrix is given by 
\begin{equation}\label{S.1.4s}
	\Delta\bar{\rho}(t)\equiv \bar{\rho}(t+\Delta t)-\bar{\rho}(t)=-i\int_{t}^{t+\Delta t}dt^{'}[\sum_m\bar{H}_{SB}^{(m)}(t^{'}),\bar{\rho}(t^{'})]. 
\end{equation}
Eq. (\ref{S.1.4s}) can be iterated by inserting for $\bar{\rho}(t^{'})$ again, which leads to
\begin{equation}\label{S.1.5}
	\Delta\bar{\rho}(t)=-i\int_{t}^{t+\Delta t}dt^{'}[\sum_m\bar{H}_{SB}^{(m)}(t^{'}),\bar{\rho}(t)]-\int_{t}^{t+\Delta t}dt^{'}\int_{t}^{t^{'}}dt^{''}[\sum_m\bar{H}_{SB}^{(m)}(t^{'}),[\sum_n\bar{H}_{SB}^{(n)}(t^{''}),\bar{\rho}(t^{''})]].
\end{equation}
Assuming that the reservoirs satisfy the Markovian condition, meaning the relaxation time $\tau_B$ of reservoirs is signifcantly shorter than the dynamical evolution timescale $\Delta t$ of the central systems, one can prove that the correlations between system and environment can be neglected \cite{Citro,Brasil}. Then the total density operator can be factorized as (Born approximation) 
\begin{equation}\label{S.1.5p}
	\bar{\rho}(t)\approx \bar{\rho}_{S}(t)\otimes \rho_B,      \quad  \rho_B=\prod_m\rho_B^{(m)}.
\end{equation}
%which amounts to neglecting higher-order system-reservoir correlations \cite{Citro}. 
Under the Markov coarse-graining condition, the reduced density matrix varies slowly within the integration window, allowing the replacement $\bar{\rho}(t^{'})\approx\bar{\rho}(t^{''})\approx\bar{\rho}(t)$. Substituting this factorization Eq. (\ref{S.1.5p}) into Eq. (\ref{S.1.5}) and tracing over the reservoir degrees of freedom yields
\begin{equation}\label{S.1.6}
	\Delta\bar{\rho}_S(t)\equiv \bar{\rho}_S(t+\Delta t) - \bar{\rho}_S(t) = \mathrm{Tr}_B\left[ \Delta\bar{\rho}(t) \right]=-\int_{t}^{t+\Delta t}dt^{'}\int_{t}^{t^{'}}dt^{''} Tr_{B}\bigg[\sum_m\bar{H}_{SB}^{(m)}(t^{'}),[\sum_n\bar{H}_{SB}^{(n)}(t^{''}),\bar{\rho}_{S}(t)\otimes\rho_B]\bigg].
\end{equation}
Because the system-reservoir coupling is linear, the term $\bar{H}_{SB}^{(m)}(t)$ contains a single (odd number of) bath creation/annihilation operator(s). For an equilibrium reservoir $Tr_B[a^{(m)}\rho_B]=Tr_B[a^{(m)\dagger}\rho_B]=0$, so the first-order term vanishes and $Tr_B[\bar{H}_{SB}^{(m)}(t),\bar{\rho}(t)]=0$. %Furthermore, the coarse-grained approximation $\bar{\rho}(t^{''})\approx\bar{\rho}(t)$ is applied, as $\bar{\rho}_S$ evolves much slower than the reservoirs. 
Substituting Eq. (\ref{S.1.3}) into Eq. (\ref{S.1.6}) we derive
\begin{equation}\label{S.1.7}
	\begin{split}
		\Delta\bar{\rho}_S(t)=-\int_{t}^{t+\Delta t}dt^{'}\int_{t}^{t^{'}}dt^{''} & \sum_{mk} \lambda_{jk}^{(m)*}\lambda_{lk}^{(m)}[\bar{c}_j^{\dagger}(t^{'})\bar{c}_l(t^{''})\bar{\rho}_S(t)f_k^{(m)}(t^{'}-t^{''}) -\bar{c}_j^{\dagger}(t^{'})\bar{\rho}_S(t)\bar{c}_l(t^{''})h_k^{(m)}(t^{'}-t^{''}) \\ & ~~~~~ -\bar{c}_l(t^{''})\bar{\rho}_S(t)\bar{c}_j^{\dagger}(t^{'})f_k^{(m)}(t^{'}-t^{''})+\bar{\rho}_S(t)\bar{c}_l(t^{''})\bar{c}_j^{\dagger}(t^{'})h_k^{(m)}(t^{'}-t^{''})] \\ + & \sum_{mk} \lambda_{jk}^{(m)}\lambda_{lk}^{(m)*}[\bar{c}_j(t^{'})\bar{c}_l^{\dagger}(t^{''})\bar{\rho}_S(t)h_k^{(m)*}(t^{'}-t^{''}) -\bar{c}_j(t^{'})\bar{\rho}_S(t)\bar{c}_l^{\dagger}(t^{''})f_k^{(m)*}(t^{'}-t^{''}) \\ & ~~~~~ -\bar{c}_l^{\dagger}(t^{''})\bar{\rho}_S(t)\bar{c}_j(t^{'})h_k^{(m)*}(t^{'}-t^{''})+\bar{\rho}_S(t)\bar{c}_l^{\dagger}(t^{''})\bar{c}_j(t^{'})f_k^{(m)*}(t^{'}-t^{''})].
	\end{split}
\end{equation}
Here the reservoir correlation functions are
\begin{equation}\label{S.1.8}
	f_k^{(m)}(t^{'}-t^{''})=Tr_B[\bar{a}_{k}^{(m)}(t^{'})\bar{a}_{k}^{(m)\dagger}(t^{''})\rho_B],~~~~~~ h_k^{(m)}(t^{'}-t^{''})=Tr_B[\bar{a}_{k}^{(m)\dagger}(t^{''})\bar{a}_{k}^{(m)}(t^{'})\rho_B].
\end{equation}
Considering that $f_k^{(m)}(\tau)$ and $h_k^{(m)}(\tau)$ rapidly decay on the timescale $\tau_B \ll \Delta t$ in the Markovian approximation, we can extend the upper limit of the time integral to infinity. Redefining the variable as $\tau = t' - t''$, Eq. (\ref{S.1.7}) simplifies to
%Considering that $f_k^{(m)}(\tau)$ and $h_k^{(m)}(\tau)$ decay in the timescale $\tau_B\ll\Delta t$ in the Markovian approximation, and redefining the variable $\tau=t^{'}-t^{''}$, then Eq. \ref{S.1.7} simplifies to
\begin{equation}\label{S.1.9}
	\begin{split}
		\Delta\bar{\rho}_S(t)=-\int_{t}^{t+\Delta t}dt^{'}\int_{0}^{\infty}d\tau & \sum_{mk} \lambda_{jk}^{(m)*}\lambda_{lk}^{(m)}[\bar{c}_j^{\dagger}(t^{'})\bar{c}_l(t^{'})\bar{\rho}_S(t)f_k^{(m)}(\tau) -\bar{c}_j^{\dagger}(t^{'})\bar{\rho}_S(t)\bar{c}_l(t^{'})h_k^{(m)}(\tau) \\ & ~~~~~ -\bar{c}_l(t^{'})\bar{\rho}_S(t)\bar{c}_j^{\dagger}(t^{'})f_k^{(m)}(\tau)+\bar{\rho}_S(t)\bar{c}_l(t^{'})\bar{c}_j^{\dagger}(t^{'})h_k^{(m)}(\tau)] \\ + & \sum_{mk} \lambda_{jk}^{(m)}\lambda_{lk}^{(m)*}[\bar{c}_j(t^{'})\bar{c}_l^{\dagger}(t^{'})\bar{\rho}_S(t)h_k^{(m)*}(\tau) -\bar{c}_j(t^{'})\bar{\rho}_S(t)\bar{c}_l^{\dagger}(t^{'})f_k^{(m)*}(\tau) \\ & ~~~~~ -\bar{c}_l^{\dagger}(t^{'})\bar{\rho}_S(t)\bar{c}_j(t^{'})h_k^{(m)*}(\tau)+\bar{\rho}_S(t)\bar{c}_l^{\dagger}(t^{'})\bar{c}_j(t^{'})f_k^{(m)*}(\tau)].
	\end{split}
\end{equation}
Taking the limit $\Delta t\rightarrow 0$, we obtain:
\begin{equation}\label{S.1.10}
	\begin{split}
		\frac{d\bar{\rho}_S(t)}{dt}= & -\sum_{m}[\bar{c}_j^{\dagger}(t)\bar{c}_l(t)\bar{\rho}_S(t)F_{jl}^{(m)} -\bar{c}_j^{\dagger}(t)\bar{\rho}_S(t)\bar{c}_l(t)H_{jl}^{(m)}  -\bar{c}_l(t)\bar{\rho}_S(t)\bar{c}_j^{\dagger}(t)F_{jl}^{(m)}+\bar{\rho}_S(t)\bar{c}_l(t)\bar{c}_j^{\dagger}(t)H_{jl}^{(m)}] \\ & -\sum_{m}[\bar{c}_j(t)\bar{c}_l^{\dagger}(t)\bar{\rho}_S(t)H_{jl}^{(m)*} -\bar{c}_j(t)\bar{\rho}_S(t)\bar{c}_l^{\dagger}(t)F_{jl}^{(m)*}  -\bar{c}_l^{\dagger}(t)\bar{\rho}_S(t)\bar{c}_j(t)H_{jl}^{(m)*}+\bar{\rho}_S(t)\bar{c}_l^{\dagger}(t)\bar{c}_j(t)F_{jl}^{(m)*}].
	\end{split}
\end{equation}
Here,
\begin{equation}\label{S.1.11}
	F_{jl}^{(m)}=\sum_{k}\lambda_{jk}^{(m)*}\lambda_{lk}^{(m)}\int_{0}^{\infty}d\tau f_k^{(m)}(\tau),~~~~~~~~ H_{jl}^{(m)}=\sum_{k}\lambda_{jk}^{(m)*}\lambda_{lk}^{(m)}\int_{0}^{\infty}d\tau h_k^{(m)}(\tau).
\end{equation}
Separating $F$ and $H$ into Hermitian and anti-Hermitian parts as $F=\bar{F}+i\tilde{F}$ and $H=\bar{H}+i\tilde{H}$, and together with Eq. (\ref{S.1.10}) we arrive at
\begin{equation}\label{S.1.12}
	\begin{split}
		\frac{d\bar{\rho}_S(t)}{dt}= & -i[\sum_{mjl}\tilde{F}_{jl}^{(m)}\bar{c}_j^{\dagger}(t)\bar{c}_l(t) +\sum_{mjl}\tilde{H}_{jl}^{(m)}\bar{c}_l(t)\bar{c}_j^{\dagger}(t),\bar{\rho}_S(t)] \\ & +\sum_{mjl}\bar{F}_{jl}^{(m)}[2\bar{c}_l(t)\bar{\rho}_S(t)\bar{c}_j^{\dagger}(t)-\{\bar{c}_j^{\dagger}(t)\bar{c}_l(t),\bar{\rho}_S(t)\}] +\sum_{mjl}\bar{H}_{jl}^{(m)}[2\bar{c}_j^{\dagger}(t)\bar{\rho}_S(t)\bar{c}_l(t)-\{\bar{c}_l(t)\bar{c}_j^{\dagger}(t),\bar{\rho}_S(t)\}].
	\end{split}
\end{equation}
Transforming back into the Schr\"{o}dinger picture via $\frac{d\bar{\rho}_S(t)}{dt}=e^{iH_St}(\frac{d\rho_S}{dt}+i[H_S,\rho_S])e^{-iH_St}$, we obtain the Lindblad equation:
\begin{equation}\label{S.1.13}
	\frac{d\rho_S}{dt}=-i[H_S+H_{Lamb},\rho_S]+ \sum_{\nu=1,2}\sum_{m,k}(2L_{\nu,mk}\rho_SL_{\nu,mk}^{\dagger}-\{L_{\nu,mk}^{\dagger}L_{\nu,mk},\rho_S\}),
\end{equation}
where the Lamb shift Hamiltonian is given by $H_{Lamb}=\sum_{mjl}(\tilde{F}_{jl}^{(m)}c_j^{\dagger}c_l+\tilde{H}_{jl}^{(m)}c_l c_j^{\dagger})$, and the Lindblad operators
\begin{equation}\label{S.1.14}
	L_{1,mk}=\sum_{j}\lambda_{jk}^{(m)}\sqrt{Re\int_{0}^{\infty}d\tau f_{k}^{(m)}(\tau)}c_{j},~~~~~~ L_{2,mk}=\sum_{j}\lambda_{jk}^{(m)*}\sqrt{Re\int_{0}^{\infty}d\tau h_{k}^{(m)}(\tau)}c_{j}^{\dagger},
\end{equation}
Hence each energy level $\epsilon_k$ in the $m$th reservoir can be described by two Lindblad dissipators $L_{1,mk}$ and $L_{2,mk}$, controlling the particle loss and gain. In general, the reservoir correlation functions can generally be expressed as:
\begin{equation}\label{S.1.15}
	f_{k}^{(m)}(\tau)=e^{-(i\epsilon_k+\frac{1}{\tau_{k}^{(m)}})\tau}(1-n_{k}^{(m)}),~~~~~~~~ h_{k}^{(m)}(\tau)=e^{-(i\epsilon_k+\frac{1}{\tau_{k}^{(m)}})\tau}n_{k}^{(m)},
\end{equation}
where $\tau_{k}^{(m)}$ represents the (qausi-)particle's lifetime, and $n_{k}^{(m)}=Tr_B[a_{k}^{(m)\dagger}a_{k}^{(m)}\rho_B]$ denotes the particle number in equilibrium. Integral over $\tau$ we get $\int_{0}^{\infty}d\tau f_{k}^{(m)}(\tau)=\frac{\tau_{k}^{(m)}-i\epsilon_k\tau_{k}^{(m)2}}{1+\epsilon_k^2\tau_{k}^{(m)2}}(1-n_{k}^{(m)})$ and $\int_{0}^{\infty}d\tau h_{k}^{(m)}(\tau)=\frac{\tau_{k}^{(m)}-i\epsilon_k\tau_{k}^{(m)2}}{1+\epsilon_k^2\tau_{k}^{(m)2}}n_{k}^{(m)}$. Hence Eq. (\ref{S.1.14}) can be simplified to
\begin{equation}\label{S.1.16}
	L_{1,mk}=\sum_{j}\lambda_{jk}^{(m)}\sqrt{\frac{\tau_{k}^{(m)}(1-n_{k}^{(m)})}{1+\epsilon_k^2\tau_{k}^{(m)2}}}c_{j},~~~~~~ L_{2,mk}=\sum_{j}\lambda_{jk}^{(m)*}\sqrt{\frac{\tau_{k}^{(m)}n_{k}^{(m)}}{1+\epsilon_k^2\tau_{k}^{(m)2}}}c_{j}^{\dagger}.
\end{equation}

In the continuous limit, it is convenient to define the local spectral density of the $m$th reservoir $J_{jl}^{(m)}=\sum_{k}\lambda_{jk}^{(m)*}\lambda_{lk}^{(m)}\delta(\omega-\epsilon_k^{m})$, then we have $\bar{F}_{jl}^{(m)}=\int d\omega J_{jl}^{(m)}(\omega)\frac{\tau_{\omega}^{(m)}(1-n_{\omega}^{(m)})}{1+\omega^2\tau_{\omega}^{(m)2}}$ and $\bar{H}_{jl}^{(m)}=\int d\omega J_{jl}^{(m)}(\omega)\frac{\tau_{\omega}^{(m)}n_{\omega}^{(m)}}{1+\omega^2\tau_{\omega}^{(m)2}}$. The Lindblad equation Eq. (\ref{S.1.13}) becomes
\begin{equation}\label{S.1.17}
	\frac{d\rho_S}{dt}=-i[H_S+H_{Lamb},\rho_S]+ \sum_{m}\bar{F}_{jl}^{(m)}(2c_l\rho_Sc_j^{\dagger}-\{c_j^{\dagger}c_l,\rho_S\})+ \sum_{m}\bar{H}_{jl}^{(m)}(2c_j^{\dagger}\rho_Sc_l-\{c_lc_j^{\dagger},\rho_S\}).
\end{equation}
Noting that $\bar{\mathbf{F}}^{(m)}$ and $\bar{\mathbf{H}}^{(m)}$ are Hermitian, positive semi-definite matrices. Their summations are diagonalized by $\mathbf{U}\sum_{m}\bar{\mathbf{F}}^{(m)}\mathbf{U}^{\dagger}=diag(x_{m,1}^2,x_{m,2}^2,\cdots$) and $\mathbf{V}\sum_m\bar{\mathbf{H}}^{(m)}\mathbf{V}^{\dagger}=diag(y_{m,1}^2,y_{m,2}^2,\cdots$), hence Eq. (\ref{S.1.17}) can be further simplified to
\begin{equation}\label{S.1.18}
	\frac{d\rho_S}{dt}=-i[H_S+H_{Lamb},\rho_S]+ \sum_{\nu=1,2}\sum_{m}(2L_{\nu,m}\rho_SL_{\nu,m}^{\dagger}-\{L_{\nu,m}^{\dagger}L_{\nu,m},\rho_S\}),
\end{equation}
with Lindblad operators
\begin{equation}\label{S.1.19}
	L_{1,m}=\sum_{j}x_{m}U_{mj}c_{j},~~~~~~ L_{2,m}=\sum_{j}y_{m}V_{mj}^{*}c_{j}^{\dagger}.
\end{equation}
By comparing Eq. (\ref{S.1.19}) with Eq. (3) in the main text, we identify $u_{mj}=x_{m}U_{mj}$ and $v_{mj}=y_{m}V_{mj}^{*}$. 

The Lamb shift term $H_{\text{Lamb}}$ originates from the imaginary part of the reservoir correlation functions and accounts for energy renormalization due to virtual transitions between the system and the reservoir. In most practical cases, this contribution is negligibly small for the following reasons. First, 
the Lamb shift appears only at second order and is typically much smaller than the coherent system energy scales. Second, we are primarily interested in dissipative dynamics, where the non-unitary evolution governed by the Lindblad terms dominates, and we are not concerned with the small energy renormalizations induced by dissipation. Third, under the Markov approximation (where reservoir modes have very short lifetimes), the rapid decay of correlation functions causes their oscillatory components to average out: $\text{Im} \int_0^\infty f_k(\tau)\, d\tau \rightarrow \tau_B^2\rightarrow 0$,
since the imaginary part cannot accumulate under fast exponential damping. In addition, system-reservoir coupling is often local in space, such that only the diagonal elements ($j = l$) of $H_{\text{Lamb}}$ contribute, leading to a weak on-site energy shift. In these circumstances, the Lamb shift has little to no observable impact on the dissipative dynamics of the system and can be safely neglected in the analysis.

Moreover, from the expressions of the gain and loss terms in Eq. (\ref{S.1.16}), we can identify the key factors that influence their strength and form.
1. The lifetime of the reservoir mode $\tau_k$.
2. The occupation number of the reservoir mode $n_k$: The loss channel scales as $L_1 \propto \sqrt{1 - n_k}$, indicating that the reservoir must be partially empty (\( n_k \neq 1 \)) to accept particles from the system. The gain channel scales as $L_2 \propto \sqrt{n_k}$, meaning that the reservoir must be partially occupied (\( n_k \neq 0 \)) in order to inject particles into the system. When \( n_k = 1 \), only gain occurs; when \( n_k = 0 \), only loss occurs; and for \( 0 < n_k < 1 \), both gain and loss processes are possible.
3. The energy detuning of the reservoir mode $\epsilon_k$: When the reservoir mode energy approaches the system energy levels (i.e., smaller \( \epsilon_k \)), the strength of gain and loss increases due to enhanced resonance.
4. The coupling strength and type between system and reservoir $\lambda_{jk}$: $\lambda_{jk}$ directly determines the strength and specific form of the Lindblad operators. In the expressions for $L_1$ and $L_2$, the summation involves $\lambda_{jk}$, which contains information about the index $j$. Therefore, it determines whether the gain or loss is on-site, bond-related, or involves lattice sites at greater distances.

%The expression of loss/gain terms are then determined by the local spectral density $J_{jl}$, (qausi-)particle lifetime $\tau$, and the distributions of hole $1-n$/particle $n$. If the reservoir is empty $n=0$, we find $\bar{H}_{jl}=0$ and hence $L_2=0$, in this case, only loss channels exist. Conversely, if the reservoir is fully occupied $n=1$, we find $\bar{F}_{jl}=0$ and hence $L_1=0$, in this case, only gain channels exist. In general, both loss and gain channels exist.

%article scatter into the Markovian reservoir, it will thermalize/dissipate rapidly and lose its initial state information.

\subsection{III. Definition of particle and energy currents}
%\section{Definition of particle and energy currents}
The steady state of the system we study (Fig. 1 in the main text) is not in equilibrium, which leads to difficulties in defining the global thermodynamic variables. To overcome this problem, it is convenient to instead study the particle and energy flows in the lead~\cite{Yamamoto}. From the first law of thermodynamics, we have
\begin{equation}\label{S.2.1}
	dU_{L(R)}=dQ_{L(R)}+dW_{L(R)},
\end{equation}
where $U_{L(R)}=\langle H_{L(R))}\rangle$ is the energy of $L(R)$-lead, $dU_{L(R)}$ and $dQ_{L(R)}$ are the energy and heat flowing into the $L(R)$-lead, and $dW_{L(R)}=\mu_{L(R)}d\langle\mathcal{N}_{L(R)}\rangle$, with $\mathcal{N}_{L(R)}$ being the particle number of the $L(R)$-lead, is the work done on the $L(R)$-lead. Typically, the voltage $V$ is defined as the difference of chemical potential, $eV=\mu_R-\mu_L$. The particle and energy currents are then defined as
\begin{equation}\label{S.2.2}
	\begin{split}
		J_{L}^{0}=-\frac{d\langle\mathcal{N}_{L}\rangle}{dt},~~~~~J_{R}^{0}=\frac{d\langle\mathcal{N}_{R}\rangle}{dt},~~~~~J_{L}^{1}=-\frac{dU_{L}}{dt},~~~~~ J_{R}^{1}=\frac{dU_{R}}{dt}.
		%J_{R}^{0}=\frac{d\mathcal{N}_{R}}{dt},~~~~~&~~~~~J_{R}^{1}=\frac{dU_{R}}{dt}.
	\end{split}
\end{equation}
Substituting into Eq. (\ref{S.2.1}), we see that the heat current is
\begin{equation}\label{S.2.3}
	J_{L}^{Q}=-\frac{dQ_{L}}{dt}=J_{L}^{1}-\mu_{L}J_{L}^{0},~~~~~~ J_{R}^{Q} = \frac{dQ_{R}}{dt}=J_{R}^{1}-\mu_{R}J_{R}^{0}.
\end{equation}
It is common to study the net current $J$ defined by
\begin{equation}\label{S.2.4}
	J^{0/1/Q}=\frac{1}{2}(J_L^{0/1/Q}+J_R^{0/1/Q}).
\end{equation}
%In the absence of dissipation, the current $\bar{J}^{0/1}$ vanish as the particle number and energy are conserved in the bulk.

\subsection{IV. Keldysh theory of open fermionic systems}
In a system coupled with the environment, one often studies the reduced density matrix $\rho$. Under
the Markovian approximation, the evolution of $\rho$ can be described by the Lindblad master equation
\begin{equation}\label{S.3.1}
	\frac{d\rho}{dt}=-i[H,\rho]+\sum_{m}(2L_{m}\rho L_{m}^{\dagger}-\{L_{m}^{\dagger}L_{m},\rho\}).
\end{equation}
Here, $H$ is the Hamiltonian (renormailzed by the environment), and $L_{m}$ are the quantum jump operators describing the coupling between the system and the environment. It is convenient to introduce the Lindblad-Keldysh partition function, $Z=Tr[\rho(t)]$,
which is always equal to one due to the normalization of the density matrix. By inserting fermionic coherent states $|\psi\rangle$, which are eigenstates of the annihilation operators: $c_{j}|\psi\rangle=\psi_j|\psi\rangle$, the partition function can be written as 
$Z=\int D[\bar{\psi}^{\pm},\psi^{\pm}]e^{i\mathcal{S}}$ \cite{Sieberer,Thompson}, where the superscript $\pm$ denotes the indices of Keldysh contour, and the action reads
\begin{equation}\label{S.3.4}
	\mathcal{S}=\int dt[\sum_{j}(\bar{\psi}_j^{+}i\partial_t\psi_j^{+}-\bar{\psi}_j^{-}i\partial_t\psi_j^{-})-i\mathcal{L}].
\end{equation}
Here, the Liouvillian $\mathcal{L}=-iH^{+}+iH^{-}-(L^{\dagger}L)^{+}-(L^{\dagger}L)^{-}+2(L)^{+}(L^{\dagger})^{-}$, 
where $H^{\pm}$ and $L^{\pm}$ satisfy:
$O^{+}=\frac{\langle\psi^{+}(t_{n+1})|O|\psi^{+}(t_{n})\rangle}{\langle\psi^{+}(t_{n+1})|\psi^{+}(t_{n})\rangle}$ and $O^{-}=\frac{\langle-\psi^{-}(t_{n})|O|-\psi^{-}(t_{n+1})\rangle}{\langle-\psi^{-}(t_{n})|-\psi^{-}(t_{n+1})\rangle}$, with $O$ representing $H$ or $L$. The description in this basis often contains redundancy. To eliminate this redundancy and facilitate calculations, a Keldysh rotation is usually performed,
%It is convenient to follow the Larkin-Ovchinnikov convention. Performing a Keldysh rotation
\begin{equation}\label{S.3.7}
	\begin{cases}
		\psi^{1}=\frac{1}{\sqrt{2}}(\psi^{+}+\psi^{-}), \\
		\psi^{2}=\frac{1}{\sqrt{2}}(\psi^{+}-\psi^{-}).
	\end{cases}~~~~~~~~
	\begin{cases}
		\bar{\psi}^{1}=\frac{1}{\sqrt{2}}(\bar{\psi}^{+}-\bar{\psi}^{-}), \\
		\bar{\psi}^{2}=\frac{1}{\sqrt{2}}(\bar{\psi}^{+}+\bar{\psi}^{-}).
	\end{cases}
\end{equation}
Hence, the Liouvillian and action can be expressed as functions of $\bar{\psi}^{1,2}$ and $\psi^{1,2}$.

Suppose the Hamiltonian is non-interacting and the Lindbladian is linear, 
\begin{equation}\label{S.3.8}
	H=\sum_{jk}h_{jk}c_{j}^{\dagger}c_{k},~~~~~~~~ L_{1,j}=\sum_{k}u_{jk}c_{k},~~~~~~~~ L_{2,j}=\sum_{k}v_{jk}c_{k}^{\dagger}.
\end{equation}
The action is quadratic and can be written in matrix form in Keldysh space:
\begin{equation}\label{S.3.9}
	\mathcal{S}=\int dt\left(
	\begin{array}{cc}
		\bar{\bm{\psi}}^{1} & \bar{\bm{\psi}}^{2} \\
	\end{array}
	\right)\left(
	\begin{array}{cc}
		i\partial_t-\mathbf{h}+i\mathbf{P}+i\mathbf{Q} & -2i(\mathbf{Q}-\mathbf{P}) \\
		0 & i\partial_t-\mathbf{h}-i\mathbf{P}-i\mathbf{Q} \\
	\end{array}
	\right)\left(
	\begin{array}{c}
		\bm{\psi}^{1} \\
		\bm{\psi}^{2} \\
	\end{array}
	\right),
\end{equation}
where the matrix element of the loss matrix $\mathbf{P}$ is $P_{jk}=\sum_{m}u_{mj}^{*}u_{mk}$, and the matrix element of the gain matrix $\mathbf{Q}$ is $Q_{jk}=\sum_{m}v_{mj}v_{mk}^{*}$. It is convenient to define the damping matrix $\mathbf{X}=\mathbf{h}-i\mathbf{P}-i\mathbf{Q}$ and the imbalance matrix $\mathbf{Y}=2(\mathbf{P}-\mathbf{Q})$. The Green's functions can be derived by inverting the action matrix 
$\mathbf{S}$, i.e.,
\begin{equation}\label{S.3.10}
	\left(
	\begin{array}{cc}
		\mathbf{g}_{jk}^{\mathcal{R}} & \mathbf{g}_{jk}^{\mathcal{K}} \\
		0 & \mathbf{g}_{jk}^{\mathcal{A}} \\
	\end{array}
	\right)=\left(
	\begin{array}{cc}
		i\partial_t-\mathbf{X} & i\mathbf{Y} \\
		0 & i\partial_t-\mathbf{X}^{\dagger} \\
	\end{array}
	\right)^{-1}.
\end{equation}
If we focus on the steady state, the Green's function depends only on the time difference $t-t^{'}$. Hence, the explicit expressions of the Green's functions in the frequency domain are
\begin{equation}\label{S.3.11}
	\mathbf{g}^{\mathcal{R}}=\frac{1}{\omega-\mathbf{X}},~~~~~~\mathbf{g}^{\mathcal{A}}=\frac{1}{\omega-\mathbf{X}^{\dagger}},~~~~~~ \mathbf{g}^{\mathcal{K}}=-\frac{1}{\omega-\mathbf{X}}i\mathbf{Y}\frac{1}{\omega-\mathbf{X}^{\dagger}}.
\end{equation}
We see that the retarded and advanced Green's functions contain only the spectral information of the damping matrix, while the Keldysh Green's function, which captures the particle distribution in the steady state, contains both the spectral information of $\mathbf{X}$ and the imbalance matrix $\mathbf{Y}$. By using the identities $\mathbf{g}^{>}=\frac{1}{2}(\mathbf{g}^{\mathcal{K}}+\mathbf{g}^{\mathcal{R}}-\mathbf{g}^{\mathcal{A}})$ and $\mathbf{g}^{<}=\frac{1}{2}(\mathbf{g}^{\mathcal{K}}-\mathbf{g}^{\mathcal{R}}+\mathbf{g}^{\mathcal{A}})$, the steady-state greater and lesser Green's functions can be derived
\begin{equation}\label{S.3.12}
	\mathbf{g}^{>}=-2i\mathbf{g}^{\mathcal{R}}\mathbf{P}\mathbf{g}^{\mathcal{A}},~~~~~~~~
	\mathbf{g}^{<}=2i\mathbf{g}^{\mathcal{R}}\mathbf{Q}\mathbf{g}^{\mathcal{A}}.
\end{equation}
We can see that the greater Green's function is directly proportional to the loss matrix $\mathbf{P}$ and the lesser Green's function is directly proportional to the gain matrix $\mathbf{Q}$.

\subsection{V. Derivation details of extended Landauer-B\"{u}ttiker formula}
Suppose the full Hamiltonian of the system shown in Fig. 1 of the main text is
\begin{equation}\label{S.4.1}
	H=H_{S}+H_{L}+H_{R}+(H_{SL}+H_{SR}+h.c.),
\end{equation}
where $H_{S}$ is the Hamiltonian of the central system, $H_{L(R)}$ are the Hamiltonians of left (right) leads, and $H_{SL(SR)}$ represents the coupling between the central system and the left (right) lead. We assume that the Markovian reservoirs only interact with the central system, and hence the Lindblad operators $L_{m}$ contain only the operators acting on the central system. The explicit form of the Hamiltonian in Eq. (\ref{S.4.1}) is
\begin{equation}\label{S.4.2}
	\begin{split}
		H_{S} =& \sum_{ij}h_{S,ij}c_{i}^{\dagger}c_{j}, ~~~~~~
		H_{L} = \sum_{k}\epsilon_{L,k}d_{L,k}^{\dagger}d_{L,k}, ~~~~~~
		H_{R} = \sum_{k}\epsilon_{R,k}d_{R,k}^{\dagger}d_{R,k}, \\
		H_{LS} =& \sum_{kj}t_{L,kj}d_{L,k}^{\dagger}c_{j}, ~~~~~~
		H_{RS} = \sum_{kj}t_{R,kj}d_{R,k}^{\dagger}c_{j}.
	\end{split}
\end{equation}
Here $c_j(c_j^{\dagger})$ is the annihilation (creation) operator on the $j$-th lattice site of the central system,
and $d_{L(R), k}(d_{L(R), k}^{\dagger})$ is the annihilation (creation) operator on the momentum $k$ of the left (right) lead. We first consider the current flowing out of the left lead, as shown in Eq. (\ref{S.2.2}), which can be calculated by
\begin{equation}\label{S.4.3}
	\begin{split}
		J_L^{0} =& -\frac{d\langle\mathcal{N}_L\rangle}{dt}=-\frac{i}{\hbar}\langle[H,\mathcal{N}_L]\rangle=\frac{1}{\hbar} \sum_{jk}(t_{L,kj}G_{SL,jk}^{<}-t_{L,kj}^{*}G_{LS,kj}^{<}), \\
		J_L^{1} =& -\frac{dU_L}{dt}=-\frac{i}{\hbar}\langle[H,H_L]\rangle=\frac{1}{\hbar} \sum_{jk}\epsilon_{L,k}(t_{L,kj}G_{SL,jk}^{<}-t_{L,kj}^{*}G_{LS,kj}^{<}),
	\end{split}
\end{equation}
where the indices $j(k)$ denote the degree of freedom of the system(leads), $G_{SL,jk}^{<}$ represents the element in the $j$-th row and $k$-th column of the matrix $\mathbf{G}_{SL}^{<}$, the particle number operator is $\mathcal{N}_{L}=\sum_{k}d_{L,k}^{\dagger}d_{L,k}$, and the energy is defined as $U_{L}=\langle H_L\rangle=\sum_{k}\epsilon_{L,k}\langle d_{L,k}^{\dagger}d_{L,k}\rangle$. To derive this equation, we used the Heisenberg equation of motion for operators, which is valid here because the leads are assumed to be uncoupled from the reservoir. In a more general scenario where the reservoir is coupled to both the central system and the leads, one would instead use the Lindblad equation for operators. In the following, we use $\mathbf{g}$ to denote the bare Green's function (i.e., the Green's function without coupling to the leads), and  $\mathbf{G}$ to denote the full Green's functions. By applying the Langreth theorem, one can obtain
\begin{equation}\label{S.4.4}
	\begin{split}
		\mathbf{G}_{LS}^{<} =& \mathbf{g}_{L}^{<}\bm{\Sigma}_{LS}^{\mathcal{A}}\mathbf{G}_{S}^{\mathcal{A}}+\mathbf{g}_{L}^{\mathcal{R}}\bm{\Sigma}_{LS}^{\mathcal{R}}\mathbf{G}_{S}^{<} +\mathbf{g}_{L}^{\mathcal{R}}\bm{\Sigma}_{L}^{<}\mathbf{G}_{LS}^{\mathcal{A}}, \\
		\mathbf{G}_{SL}^{<} =& \mathbf{G}_{S}^{<}\bm{\Sigma}_{SL}^{\mathcal{A}}\mathbf{g}_{L}^{\mathcal{A}}+\mathbf{G}_{S}^{\mathcal{R}}\bm{\Sigma}_{SL}^{\mathcal{R}}\mathbf{g}_{L}^{<} +\mathbf{G}_{SL}^{\mathcal{R}}\bm{\Sigma}_{L}^{<}\mathbf{g}_{L}^{\mathcal{A}}.
	\end{split}
\end{equation}
and 
\begin{equation}\label{S.4.5}
	\begin{split}
		\mathbf{G}_{S}^{<} =& \mathbf{g}_{S}^{<}+(\mathbf{g}_{S}^{<}\bm{\Sigma}_{SL}^{\mathcal{A}}\mathbf{G}_{LS}^{\mathcal{A}} +\mathbf{g}_{S}^{\mathcal{R}}\bm{\Sigma}_{SL}^{\mathcal{R}}\mathbf{G}_{LS}^{<}+\mathbf{g}_{S}^{\mathcal{R}}\bm{\Sigma}_{SL}^{<}\mathbf{G}_{LS}^{\mathcal{A}}+L\leftrightarrow R), \\ \mathbf{G}_{S}^{>} =& \mathbf{g}_{S}^{>}+(\mathbf{g}_{S}^{>}\bm{\Sigma}_{SL}^{\mathcal{A}}\mathbf{G}_{LS}^{\mathcal{A}} +\mathbf{g}_{S}^{\mathcal{R}}\bm{\Sigma}_{SL}^{\mathcal{R}}\mathbf{G}_{LS}^{>}+\mathbf{g}_{S}^{\mathcal{R}}\bm{\Sigma}_{SL}^{>}\mathbf{G}_{LS}^{\mathcal{A}}+L\leftrightarrow R).
	\end{split}
\end{equation}
Here, $\bm{\Sigma}_{L}^{<}$ vanishes for non-interacting, non-dissipating leads. In equilibrium, the bare Green's functions of the left lead are
\begin{equation}\label{S.4.6}
	\begin{split}
		g_{L,k}^{<} = 2\pi if_{L}\delta(\omega-\epsilon_{L,k}), ~~~~~~~
		g_{L,k}^{>} = 2\pi i(f_{L}-1)\delta(\omega-\epsilon_{L,k}),
	\end{split}
\end{equation}
where $f_{L}=\frac{1}{e^{(\omega-\mu_{L})/K_{B}T_{L}}+1}$ is the Fermi distribution. Transforming into the frequency domain, $G(t-t^{'})=\int\frac{d\omega}{2\pi}e^{-i\frac{\omega(t-t^{'})}{\hbar}}G(\omega)$, and substituting Eq. (\ref{S.4.4}) and (\ref{S.4.6}) into Eq. (\ref{S.4.3}), we get
\begin{equation}\label{S.4.7}
	J_L^{\lambda}=\frac{i}{\hbar}\int\frac{d\omega}{2\pi}\omega^{\lambda} Tr[f_{L}\bm{\Gamma}_{L}(\mathbf{G}_{S}^{\mathcal{R}}-\mathbf{G}_{S}^{\mathcal{A}})+\bm{\Gamma}_{L}\mathbf{G}_{S}^{<}].
\end{equation}
Here, $\lambda=0, 1$, and the spectral density is $\bm{\Gamma}_{L}=i(\tilde{\bm{\Sigma}}_{L}^{\mathcal{R}}-\tilde{\bm{\Sigma}}_{L}^{\mathcal{A}})$, with the leads-induced self-energies $\tilde{\bm{\Sigma}}_{L}^{\mathcal{R}}=\bm{\Sigma}_{SL}^{\mathcal{R}}\mathbf{g}_{L}^{\mathcal{R}}\bm{\Sigma}_{LS}^{\mathcal{R}}$, and $\tilde{\bm{\Sigma}}_{L}^{\mathcal{A}}=\bm{\Sigma}_{SL}^{\mathcal{A}}\mathbf{g}_{L}^{\mathcal{A}}\bm{\Sigma}_{LS}^{\mathcal{A}}$. The expression for $J_{R}^{\lambda}$ can be derived in the same way. Substituting $J_{L}^{\lambda}$ and $J_{R}^{\lambda}$ into Eq. (\ref{S.2.4}), the current $J^{\lambda}$ can be written as
\begin{equation}\label{S.4.8}
	J^{\lambda}=\frac{i}{2\hbar}\int\frac{d\omega}{2\pi}\omega^{\lambda} Tr[(f_{L}\bm{\Gamma}_{L}-f_{R}\bm{\Gamma}_{R})(\mathbf{G}_{S}^{\mathcal{R}}-\mathbf{G}_{S}^{\mathcal{A}})+(\bm{\Gamma}_{L}-\bm{\Gamma}_{R})\mathbf{G}_{S}^{<}].
\end{equation}
This formula applies to a general many-body Hamiltonian and Lindbladian. It can be seen that all the information about dissipation is contained in the Green's function $\mathbf{G}_S$. In the absence of dissipation, Eq. (\ref{S.4.8}) reduces to the Meir-Wingreen formula \cite{Meir}. Additionally, this formulation inherently accounts for the system-leads correlations through the lead-induced self-energies, $\tilde{\bm{\Sigma}}_{L/R}$, which dress the central system's Green's function. Therefore, the correlations between the system and the leads are included in the steady-state transport. By multiplying $\omega-\mathbf{X}$ from the left of $\mathbf{G}_{S}^{<}(\mathbf{G}_{S}^{>})$ in Eq. (\ref{S.4.5}) and using the identities in Eq. (\ref{S.3.12}), we obtain
\begin{equation}\label{S.4.9}
	\begin{split}
		\mathbf{G}_{S}^{<} =& \mathbf{G}_{S}^{\mathcal{R}}(\bm{\Sigma}_{SL}^{\mathcal{R}}\mathbf{g}_{L}^{<}\bm{\Sigma}_{LS}^{\mathcal{A}} +\bm{\Sigma}_{SR}^{\mathcal{R}}\mathbf{g}_{\mathcal{R}}^{<}\bm{\Sigma}_{RS}^{\mathcal{A}}+2i\mathbf{Q})\mathbf{G}_{S}^{\mathcal{A}}, \\ \mathbf{G}_{S}^{>} =& \mathbf{G}_{S}^{\mathcal{R}}(\bm{\Sigma}_{SL}^{\mathcal{R}}\mathbf{g}_{L}^{>}\bm{\Sigma}_{LS}^{\mathcal{A}} +\bm{\Sigma}_{SR}^{\mathcal{R}}\mathbf{g}_{\mathcal{R}}^{>}\bm{\Sigma}_{RS}^{\mathcal{A}}-2i\mathbf{P})\mathbf{G}_{S}^{\mathcal{A}},
	\end{split}
\end{equation}
where we assume no dissipation in the leads, such that $\bm{\Sigma}_{SL}^{\gtrless}=\bm{\Sigma}_{SR}^{\gtrless}=0$. Then, noting that $\bm{\Sigma}_{SL(R)}^{\mathcal{R}}\mathbf{g}_{L(R)}^{<}\bm{\Sigma}_{L(R)S}^{\mathcal{A}}=if_{L(R)}\bm{\Gamma}_{L(R)}$ and $\bm{\Sigma}_{SL(R)}^{\mathcal{R}}\mathbf{g}_{L(R)}^{>}\bm{\Sigma}_{L(R)S}^{\mathcal{A}}=i(f_{L(R)}-1)\bm{\Gamma}_{L(R)}$, we finally obtain
\begin{equation}\label{S.4.10}
	\begin{split}
		\mathbf{G}_{S}^{<} =& i\mathbf{G}_{S}^{\mathcal{R}}(f_L\bm{\Gamma}_L+f_R\bm{\Gamma}_R+2\mathbf{Q})\mathbf{G}_{S}^{\mathcal{A}}, \\ \mathbf{G}_{S}^{>} =& i\mathbf{G}_{S}^{\mathcal{R}}(f_L\bm{\Gamma}_L+f_R\bm{\Gamma}_R-\bm{\Gamma}_L-\bm{\Gamma}_R-2\mathbf{P})\mathbf{G}_{S}^{\mathcal{A}}.
	\end{split}
\end{equation}
where $\mathbf{G}_S^{\mathcal{R}}=(\mathbf{G}_S^{\mathcal{A}})^{\dagger}=[(\mathbf{g}_S^{\mathcal{R}})^{-1}-\tilde{\bm{\Sigma}}_{L}^{\mathcal{R}}-\tilde{\bm{\Sigma}}_{R}^{\mathcal{R}}]^{-1}$,
and $(\mathbf{g}_S^{\mathcal{R}})^{-1}=\omega-\mathbf{h}_S+i\mathbf{P}+i\mathbf{Q}$. Substituting Eq. (\ref{S.4.10}) into  Eq. (\ref{S.4.8}) and using the identity $\mathbf{G}_S^{\mathcal{R}}-\mathbf{G}_S^{\mathcal{A}}=\mathbf{G}_S^{>}-\mathbf{G}_S^{<}$, the current can be written in a simple form
\begin{equation}\label{S.4.12}
	J^{\lambda}=J_{LR}^{\lambda}+J_{LS}^{\lambda}-J_{RS}^{\lambda},
\end{equation}
with
\begin{equation}\label{S.4.13}
	\begin{split}
		J_{LR}^{\lambda} =& \int\frac{d\omega}{2h}(f_L-f_R)\omega^{\lambda}Tr[\bm{\Gamma}_{L}\mathbf{G}_{S}^{\mathcal{R}}\bm{\Gamma}_{R}\mathbf{G}_{S}^{\mathcal{A}} +\bm{\Gamma}_{R}\mathbf{G}_{S}^{\mathcal{R}}\bm{\Gamma}_{L}\mathbf{G}_{S}^{\mathcal{A}}], \\ J_{LS}^{\lambda} =& \int\frac{d\omega}{h}f_{L}\omega^{\lambda}Tr[\bm{\Gamma}_{L}\mathbf{G}_{S}^{\mathcal{R}}\mathbf{P}\mathbf{G}_{S}^{\mathcal{A}}] +\int\frac{d\omega}{h}(f_{L}-1)\omega^{\lambda}Tr[\bm{\Gamma}_{L}\mathbf{G}_{S}^{\mathcal{R}}\mathbf{Q}\mathbf{G}_{S}^{\mathcal{A}}], \\ J_{RS}^{\lambda} =& \int\frac{d\omega}{h}f_{R}\omega^{\lambda}Tr[\bm{\Gamma}_{R}\mathbf{G}_{S}^{\mathcal{R}}\mathbf{P}\mathbf{G}_{S}^{\mathcal{A}}] +\int\frac{d\omega}{h}(f_{R}-1)\omega^{\lambda}Tr[\bm{\Gamma}_{R}\mathbf{G}_{S}^{\mathcal{R}}\mathbf{Q}\mathbf{G}_{S}^{\mathcal{A}}].
	\end{split}
\end{equation}
Here, the Landauer term $J_{LR}^{\lambda}$ describes the current flow from the left lead to the right lead directly, and $J_{LS}^{\lambda}-J_{RS}^{\lambda}$ represents the current flow from the left lead to the right lead through the reservoirs.

Though Eq. (\ref{S.4.12}) gives clear physical origin for the steady-state current, it is more convenient to separate the current into
\begin{equation}\label{S.4.14}
	J^{\lambda}=J_0^{\lambda}+\delta J_1^{\lambda},
\end{equation}
where the reference current $J_0^{\lambda}$ represents the net current if both leads are at the same chemical potential and temperature. The response current $\delta J_1^{\lambda}$ captures the change in current induced by a chemical potential and temperature gradient. Suppose the chemical potential and temperature gradient are applied to the left lead, i.e., $\mu_R=\mu$, $T_{R}=T$, $f_{R}=f$, and $\mu_{L}=\mu_{R}+\delta\mu$, $T_{L}=T_{R}+\delta T$, $f_{L}=f_{R}+\delta f$. Then, we can derive
\begin{equation}\label{S.4.15}
	%\begin{split}
	J_0^{\lambda} = \int\frac{d\omega}{h}\omega^{\lambda}f(\omega)\tau_{0,l}(\omega) +\int\frac{d\omega}{h}\omega^{\lambda}(f(\omega)-1)\tau_{0,g}(\omega), ~~~~~~~~~~~~~
	\delta J_1^{\lambda} = \int\frac{d\omega}{h}\omega^{\lambda}\delta f(\omega)\tau_1(\omega),
	%\end{split}
\end{equation}
where the transmission function from the loss channels is $\tau_{0,l}(\omega)=Tr[(\bm{\Gamma}_{L}-\bm{\Gamma}_R)\mathbf{G}_{S}^{\mathcal{R}}\mathbf{P}\mathbf{G}_{S}^{\mathcal{A}}]$, the transmission function from the gain channels is $\tau_{0,g}(\omega)=Tr[(\bm{\Gamma}_{L}-\bm{\Gamma}_R)\mathbf{G}_{S}^{\mathcal{R}}\mathbf{Q}\mathbf{G}_{S}^{\mathcal{A}}]$, and the transmission function $\tau_1$ is $\tau_1(\omega)=Tr[\frac{1}{2}\bm{\Gamma}_{L}\mathbf{G}_{S}^{\mathcal{R}}\bm{\Gamma}_{R}\mathbf{G}_{S}^{\mathcal{A}}+ \frac{1}{2}\bm{\Gamma}_{R}\mathbf{G}_{S}^{\mathcal{R}}\bm{\Gamma}_{L}\mathbf{G}_{S}^{\mathcal{A}} +\bm{\Gamma}_{L}\mathbf{G}_{S}^{\mathcal{R}}(\mathbf{P}+\mathbf{Q})\mathbf{G}_{S}^{\mathcal{A}}]$.

\section{VI. Steady-state currents in bosonic systems with gain and loss}
This section generalizes the extended Landauer-B\"{u}ttiker formula derived for fermions to bosonic systems. The fundamental differences in quantum statistics, specifically stimulated emission for bosons and Pauli blocking for fermions, lead to essential distinctions in their steady-state transport properties.

We assume the system Hamiltonian and Lindblad operators have the same form as in the fermionic case [Eq. (\ref{S.3.8})], but with the creation and annihilation operators $c_j$, $c_j^{\dagger}$ satisfying bosonic commutation relations. The Lindblad-Keldysh action reads \cite{Thompson}
\begin{equation}
	S=\int dt\left(
	\begin{array}{cc}
		\bar{\bm{\phi}}^{c} & \bar{\bm{\phi}}^{q} \\
	\end{array}
	\right)\left(
	\begin{array}{cc}
		0 & i\partial_t-\mathbf{h}-i\mathbf{P}+i\mathbf{Q} \\
		i\partial_t-\mathbf{h}+i\mathbf{P}-i\mathbf{Q} & 2i(\mathbf{Q}+\mathbf{P}) \\
	\end{array}
	\right)\left(
	\begin{array}{c}
		\bm{\phi}^{c} \\
		\bm{\phi}^{q} \\
	\end{array}
	\right),
\end{equation}
where $\bm{\phi}^{c}=\frac{\bm{\phi}^{+}+\bm{\phi}^{-}}{\sqrt{2}}$, $\bm{\phi}^{q}=\frac{\bm{\phi}^{+}-\bm{\phi}^{-}}{\sqrt{2}}$, and the fields $\bm{\phi}^{\pm}$ correspond to the forward/backward Keldysh time contours. In the steady state, the system's Green's function are
\begin{equation}
	\left(
	\begin{array}{cc}
		\mathbf{g}^{\mathcal{K}} & \mathbf{g}^{\mathcal{R}} \\
		\mathbf{g}^{\mathcal{A}} & 0 \\
	\end{array}
	\right)=\left(
	\begin{array}{cc}
		0 & \omega-\mathbf{X}^{\dagger} \\
		\omega-\mathbf{X} & i\mathbf{Y} \\
	\end{array}
	\right)^{-1}=\left(
	\begin{array}{cc}
		-\frac{1}{\omega-\mathbf{X}}i\mathbf{Y}\frac{1}{\omega-\mathbf{X}^{\dagger}} & \frac{1}{\omega-\mathbf{X}} \\
		\frac{1}{\omega-\mathbf{X}^{\dagger}} & 0 \\
	\end{array}
	\right),
\end{equation}
where $\mathbf{X}=\mathbf{h}-i\mathbf{P}+i\mathbf{Q}$ and $\mathbf{Y}=2(\mathbf{P}+\mathbf{Q})$. Compared to the fermionic case [Eq. (\ref{S.3.11})], the expressions for the Green's functions are identical. However, the sign of gain matrix $\mathbf{Q}$ flips in $\mathbf{X}$ and $\mathbf{Y}$. The resulting steady-state greater and lesser Green's functions are 
\begin{equation}
	\mathbf{g}^{>}=-2i\mathbf{g}^{\mathcal{R}}\mathbf{P}\mathbf{g}^{\mathcal{A}},~~~~~~~~
	\mathbf{g}^{<}=-2i\mathbf{g}^{\mathcal{R}}\mathbf{Q}\mathbf{g}^{\mathcal{A}}.
\end{equation}

Now consider the system connected to leads at both ends. The currents are defined as
\begin{equation}\label{bosoncurrentX}
	\begin{split}
		J_L^{0} =& -\frac{d\langle\mathcal{N}_L\rangle}{dt}=-\frac{i}{\hbar}\langle[H,\mathcal{N}_L]\rangle=-\frac{1}{\hbar} \sum_{jk}(t_{L,kj}G_{SL,jk}^{<}-t_{L,kj}^{*}G_{LS,kj}^{<}), \\
		J_L^{1} =& -\frac{dU_L}{dt}=-\frac{i}{\hbar}\langle[H,H_L]\rangle=-\frac{1}{\hbar} \sum_{jk}\epsilon_{L,k}(t_{L,kj}G_{SL,jk}^{<}-t_{L,kj}^{*}G_{LS,kj}^{<}).
	\end{split}
\end{equation}
Compared with the current expression for fermions in Eq. (\ref{S.4.3}), the final equality here contains an additional minus sign, which originates from the commutation relations of bosons, in contrast to those of fermions.
At the same time, the bare Green's functions of the leads also differ from the fermionic system by a minus sign. For bosons, their expressions are:
\begin{equation}
	g_{L,k}^{<}=-2\pi i\tilde{f}_{L}\delta(\omega-\omega_{L,k}),~~~~~~~~~~
	g_{L,k}^{>}=-2\pi i(\tilde{f}_{L}+1)\delta(\omega-\omega_{L,k}),
\end{equation}
where the bosonic distribution function is $\tilde{f}_{\alpha}=\frac{1}{e^{(\omega-\mu_\alpha)/k_BT_{\alpha}}-1}$ ($\alpha\in \{L,R\}$). Applying the Langreth theorem, the generalized Meir-Wingreen formula [Eq. (\ref{S.4.8})] should be modified to:
\begin{equation}\label{Current2boson}
	J^{\lambda}=\frac{i}{2\hbar}\int\frac{d\omega}{2\pi}\omega^{\lambda} Tr[(\tilde{f}_{L}\bm{\Gamma}_{L}-\tilde{f}_{R}\bm{\Gamma}_{R})(\mathbf{G}_{S}^{\mathcal{R}}-\mathbf{G}_{S}^{\mathcal{A}}) -(\bm{\Gamma}_{L}-\bm{\Gamma}_{R})\mathbf{G}_{S}^{<}].
\end{equation}
We note that Ref.~\cite{TYamamoto} has also employed the Meir-Wingreen formula to study bosonic systems.

In the non-interacting limit, for bosonic systems, Eq. (\ref{S.4.10}) should be replaced by
\begin{equation}
	\begin{split}
		\mathbf{G}_{S}^{<} =& i\mathbf{G}_{S}^{\mathcal{R}}(-\tilde{f}_L\bm{\Gamma}_L-\tilde{f}_R\bm{\Gamma}_R-2\mathbf{Q})\mathbf{G}_{S}^{\mathcal{A}}, \\ \mathbf{G}_{S}^{>} =& i\mathbf{G}_{S}^{\mathcal{R}}(-\tilde{f}_L\bm{\Gamma}_L-\tilde{f}_R\bm{\Gamma}_R-\bm{\Gamma}_L-\bm{\Gamma}_R-2\mathbf{P})\mathbf{G}_{S}^{\mathcal{A}}.
	\end{split}
\end{equation}
Substituting these into Eq. (\ref{Current2boson}), the steady-state current can be written as
$J^{\lambda}=J_{LR}^{\lambda}+J_{LS}^{\lambda}-J_{RS}^{\lambda}$,
with
\begin{equation}
	\begin{split}
		J_{LR}^{\lambda} =& \int\frac{d\omega}{2h}(\tilde{f}_L-\tilde{f}_R)\omega^{\lambda}Tr[\bm{\Gamma}_{L}\mathbf{G}_{S}^{\mathcal{R}}\bm{\Gamma}_{R}\mathbf{G}_{S}^{\mathcal{A}} +\bm{\Gamma}_{R}\mathbf{G}_{S}^{\mathcal{R}}\bm{\Gamma}_{L}\mathbf{G}_{S}^{\mathcal{A}}], \\ J_{LS}^{\lambda} =& \int\frac{d\omega}{h}\tilde{f}_{L}\omega^{\lambda}Tr[\bm{\Gamma}_{L}\mathbf{G}_{S}^{\mathcal{R}}\mathbf{P}\mathbf{G}_{S}^{\mathcal{A}}] -\int\frac{d\omega}{h}(\tilde{f}_{L}+1)\omega^{\lambda}Tr[\bm{\Gamma}_{L}\mathbf{G}_{S}^{\mathcal{R}}\mathbf{Q}\mathbf{G}_{S}^{\mathcal{A}}], \\ J_{RS}^{\lambda} =& \int\frac{d\omega}{h}\tilde{f}_{R}\omega^{\lambda}Tr[\bm{\Gamma}_{R}\mathbf{G}_{S}^{\mathcal{R}}\mathbf{P}\mathbf{G}_{S}^{\mathcal{A}}] -\int\frac{d\omega}{h}(\tilde{f}_{R}+1)\omega^{\lambda}Tr[\bm{\Gamma}_{R}\mathbf{G}_{S}^{\mathcal{R}}\mathbf{Q}\mathbf{G}_{S}^{\mathcal{A}}].
	\end{split}
\end{equation}
It is evident that the Landauer term $J_{LR}^{\lambda}$ remains identical to that in fermionic systems. The fundamental distinction arises from the gain-induced current: for bosons, it is proportional to $1+\tilde{f}$, reflecting the bosonic enhancement effect, whereas for fermions, the corresponding factor is $1-f$, manifesting the Pauli blocking effect.

Analogous to the fermionic case, the total current can also be decomposed as $J=J_0+\delta J_1$, with
\begin{equation}
	J_0^{\lambda}=\int\frac{d\omega}{h}\omega^{\lambda}\tilde{f}(\omega)\tau_{0,l}(\omega) -\int\frac{d\omega}{h}\omega^{\lambda}(\tilde{f}+1)\tau_{0,g}(\omega),~~~~~~~~
	\delta J_1^{\lambda}=\int\frac{d\omega}{h}\omega^{\lambda}\delta\tilde{f}(\omega)\tau_1(\omega),
\end{equation}
where the transmission functions are defined as: $\tau_{0,l}(\omega)=Tr[(\bm{\Gamma}_{L}-\bm{\Gamma}_R)\mathbf{G}_{S}^{\mathcal{R}}\mathbf{P}\mathbf{G}_{S}^{\mathcal{A}}]$,  $\tau_{0,g}(\omega)=Tr[(\bm{\Gamma}_{L}-\bm{\Gamma}_R)\mathbf{G}_{S}^{\mathcal{R}}\mathbf{Q}\mathbf{G}_{S}^{\mathcal{A}}]$, and $\tau_1(\omega)=Tr[\frac{1}{2}\bm{\Gamma}_{L}\mathbf{G}_{S}^{\mathcal{R}}\bm{\Gamma}_{R}\mathbf{G}_{S}^{\mathcal{A}}+ \frac{1}{2}\bm{\Gamma}_{R}\mathbf{G}_{S}^{\mathcal{R}}\bm{\Gamma}_{L}\mathbf{G}_{S}^{\mathcal{A}} +\bm{\Gamma}_{L}\mathbf{G}_{S}^{\mathcal{R}}(\mathbf{P}-\mathbf{Q})\mathbf{G}_{S}^{\mathcal{A}}]$.
In summary, the generalization to bosons primarily involves replacing the Fermi distribution $f$ with the Bose distribution 
$\tilde{f}$ and noting the sign flip preceding the gain matrix.

\section{VII. Effect of inversion symmetry in gain or loss terms on the current: a two-site example}
In the main text, we have mentioned that when there is no chemical potential or temperature difference between the leads, the breaking of inversion symmetry in the gain or loss terms, or in the system, can lead to the generation of current. Consider a two-site example with Hamiltonian $H_{S}=-(c_{1}^{\dagger}c_{2}+c_{2}^{\dagger}c_{1})$. The local on-site monitoring is described by Lindblad operators $L_{1}=\sqrt{\gamma_{1}}c_1$ and $L_{2}=\sqrt{\gamma_{2}}c_2$. The leads self-energies $(\tilde{\mathbf{\Sigma}}_{L}^{\mathcal{R}})_{11}=(\tilde{\mathbf{\Sigma}}_{R}^{\mathcal{R}})_{NN}=-\frac{i\Gamma}{2}$  are constants in the wide-band limit. The particle current in Eq. (\ref{S.4.14}) is $J_0^0=\Gamma\int\frac{d\omega}{h}f(\omega)\tau_{0,l}(\omega)$ with
\begin{equation}\label{S.5.1}
	\tau_{0,l}(\omega)=\frac{\gamma_1(|\omega+i\gamma_{2}+\frac{i\Gamma}{2}|^2-1) -\gamma_2(|\omega+i\gamma_{1}+\frac{i\Gamma}{2}|^2-1)} {|(\omega+i\gamma_{1}+\frac{i\Gamma}{2})(\omega+i\gamma_{2}+\frac{i\Gamma}{2})-1|^2}.
\end{equation}
Now consider the situation where the monitoring strength is very small, i.e., $\gamma_1,\gamma_2\ll\omega$. Then, Eq. (\ref{S.5.1}) can be simplified to
\begin{equation}\label{S.5.2}
	\tau_{0,l}(\omega)\approx(\gamma_1-\gamma_2)\frac{|\omega+\frac{i\Gamma}{2}|^2-1 }{|(\omega+\frac{i\Gamma}{2})^2-1|^2}.
\end{equation}
As a result, $J_0^0\propto\gamma_1-\gamma_2$ for weak $\gamma_{1}$ and $\gamma_2$, and in the presence of inversion symmetry, where $\gamma_1=\gamma_2$, the current $J_0^0$ vanishes, as shown in Fig. 2(b2) in the main text.

\begin{figure}[t]
	\centering
	\includegraphics[width=0.6\linewidth]{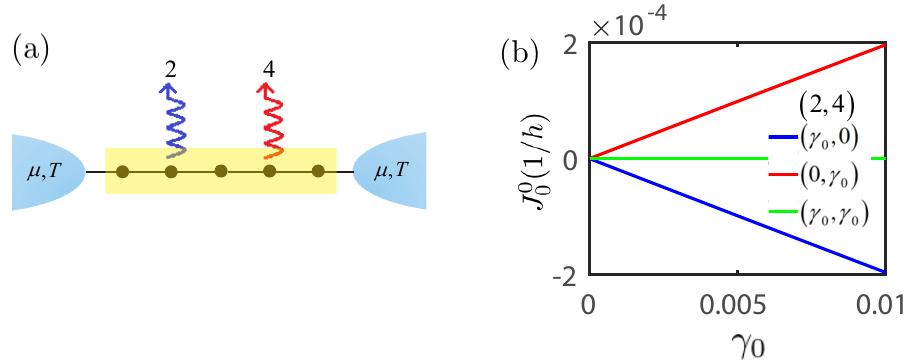}\\
	\caption{(a) Scheme of a five-site system with local monitoring. (b) The current $J_0^{0}$ as a function of monitoring strength $\gamma_0$. The blue and red lines represent monitoring at the second and fourth site, respectively; the green line corresponds to monitoring at the second and fourth sites situationally.}\label{FigS2}
\end{figure}

Numerically, we checked that the results can be generalized to the model with more sites. Fig. \ref{FigS2} shows an example with five sites. Similar to the $N=2$ case, the current $J_0^0$ is proportional to $\gamma_0$ for small monitoring at the second or fourth site. However, the current vanishes when measuring at the second and fourth sites due to inversion symmetry. It can be verified that adding gain or loss terms at more sites leads to the same conclusion: the presence or absence of inversion symmetry causes the current to vanish or emerge.

\section{VIII. Disorder-induced current generation}
In this section,  we first analyze the upper and lower limits of the linear shape of the envelope curve of the current as a function of disorder strength $V$ in Fig. 2(c) in the main text, and further explain how, as $V$ increases, the upper and lower limits of the envelope curve change from being linearly dependent on $V$ to being independent of $V$. To analyze these effects clearly, we first consider a two-site example. The Hamiltonian of the model is
$H_{S}=-(c_{1}^{\dagger}c_{2}+h.c.)+\sum_{j=1}^{2}V_jc_{j}^{\dagger}c_{j}$, where the potential $V_j$ is randomly distributed in $[-V,V]$. The on-site loss terms are $L_{j}=\sqrt{\gamma}c_{j}$, with $j=1,2$. When $f_L=f_R=f$, the current $J^{0}$ can be represented as $J^0=\Gamma\int\frac{d\omega}{h}f(\omega)\tau_{0,l}(\omega)$ with  $\tau_{0,l}=Tr[(\bm{\Gamma}_L+\bm{\Gamma}_R)\mathbf{G}_S^{\mathcal{R}}\mathbf{P}\mathbf{G}_S^{\mathcal{R}}]$ and $\mathbf{G}_S^{\mathcal{R}}=\frac{1}{\omega-\mathbf{h}_S+i\mathbf{P}}$. Here $\bm{\Gamma}_L=[\Gamma,0;0,0]$, $\bm{\Gamma}_R=[0,0;0,\Gamma]$, $\mathbf{h}_S=[V_1,-1;-1,V_2]$ and $\mathbf{P}=[\gamma,0;0,\gamma]$. After a straightforward calculation, we derive
%transmission function (details are in section VI of the Supplementary Material)
\begin{equation}\label{R.7}
	\tau_{0,l}(\omega)=\gamma\frac{\bar{V}_2^2-\bar{V}_1^2} {(\bar{V}_1\bar{V}_2-\bar{\gamma}^2-1)^2+\bar{\gamma}^2(\bar{V}_1+\bar{V}_2)^2},
\end{equation}
where the shifted potentials $\bar{V}_1=\omega-V_1$ and  $\bar{V}_2=\omega-V_2$ are randomly distributed in $[\omega-V,\omega+V]$, and $\bar{\gamma}=\gamma+\frac{\Gamma}{2}$. The envelope curve of $J^0$ is determined by the maximum value of $\tau_{0,l}$, which is difficult to calculate in general. However, it can be derived in the weak disorder limit (neglecting the $V_2^2-V_1^2$ term in the numerator, the denominator satisfies $(\bar{V_1}+\bar{V_2})^2\sim 4\omega^2$, $\bar{V_1}\bar{V_2}\sim\omega^2$), the current can be approximated by
\begin{equation}\label{R.8}
	J^0=\gamma\Gamma\int\frac{d\omega}{h}f(\omega)\frac{2\omega(V_1-V_2)} {4\omega^2\bar{\gamma}^2+(\omega^2-\bar{\gamma}^2-1)^2}.
\end{equation}
Hence if the potential is symmetric $V_1=V_2$, the current vanishes $J^0=0$. Conversely, if the potential is the ``most'' asymmetric $V_1=-V_2=\pm V$, the current reaches its maximum. The envelope curve is then $J^0\sim\pm cV$ with $c=\int\frac{d\omega}{h}f(\omega)\frac{4\omega\gamma\Gamma} {4\omega^2\bar{\gamma}^2+(\omega^2-\bar{\gamma}^2-1)^2}$. This represents the upper and lower limits of the linear envelope, and it can be seen that as the loss strength 
$\gamma\rightarrow 0$, $c$ becomes $0$, and $J^0$ also becomes $0$.

\begin{figure}[htbp]
	\centering
	\includegraphics[width=0.6\linewidth]{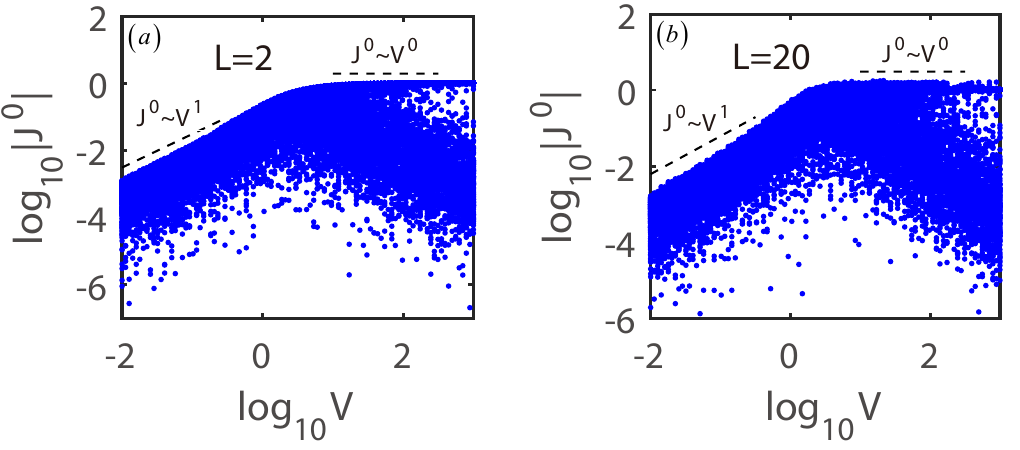}\\
	\caption{The current $J^0$ as a function of disorder strength $V$ in the zero temperature limit. The chemical potentials $\mu_L=\mu_R=0.1$. The system sizes are (a) $L=2$ and (b) $L=20$.}\label{FigSSS}
\end{figure}

In the strong disorder limit, the hopping between two sites can be neglected ($V_1, V_2\gg 1$, where $1$ is the hopping strength we chose). We can view the left lead couples to the first site, and the right lead couples to the second site. In this case, 
Particles can only effectively propagate at the resonance points $\omega\sim V_1$ and $\omega\sim V_2$, and the response rapidly decreases away from these resonance points. Therefore, the main contribution to the transmission function comes from these frequency points, which can be approximated as two peaks of a Lorentzian function. Substitute $\omega=V_1$ and $\omega=V_2$ into Eq. (\ref{R.7}) (note: $\bar{V}_1=\omega-V_1$ and  $\bar{V}_2=\omega-V_2$), and obtain the coefficients in front of $\delta(\omega-V_1)$ and $\delta(\omega-V_2)$ respectively. Combining them gives
\begin{equation}\label{R.81}
	\tau_{0,l}(\omega)\approx\gamma\frac{(V_1-V_2)^2}{(1+\bar{\gamma}^2)^2+\bar{\gamma}^2(V_1-V_2)^2} [\delta(\omega-V_1)-\delta(\omega-V_2)].
\end{equation}
Then the current is
\begin{equation}\label{R.82}
	J^0=\frac{\gamma\Gamma}{h}[f(V_1)-f(V_2)]\frac{(V_1-V_2)^2}{(1+\bar{\gamma}^2)^2+\bar{\gamma}^2(V_1-V_2)^2}.
\end{equation}
Since $|f(V_1)-f(V_2)|\leq 1$ and $\frac{(V_1-V_2)^2}{(1+\bar{\gamma}^2)^2+\bar{\gamma}^2(V_1-V_2)^2}\leq \frac{4V^2}{(1+\bar{\gamma}^2)^2+4\bar{\gamma}^2V^2}\leq\frac{1}{\bar{\gamma}^2}$, in the strong disorder limit, the maximum value of $J^0$ is finite and will not increase indefinitely, and it is no longer dependent on the value of 
$V$. In Fig. \ref{FigSSS}(a) we show the norm of the current $|J^0|$ as a function of disorder strength $V$ in the logarithmic coordinate. We find $J^0\sim V$ for small disorder strength, and $J^0\sim V^0$ for large disorder strength. Hence as disorder strength $V$ increases, the current increases linearly and then reach its saturation. Though our analysis are based on the two site example, the behavior is similar for large system size from numerical calculation, as shown in Fig. \ref{FigSSS}(b). 

\begin{figure}[htbp]
	\centering
	\includegraphics[width=1\linewidth]{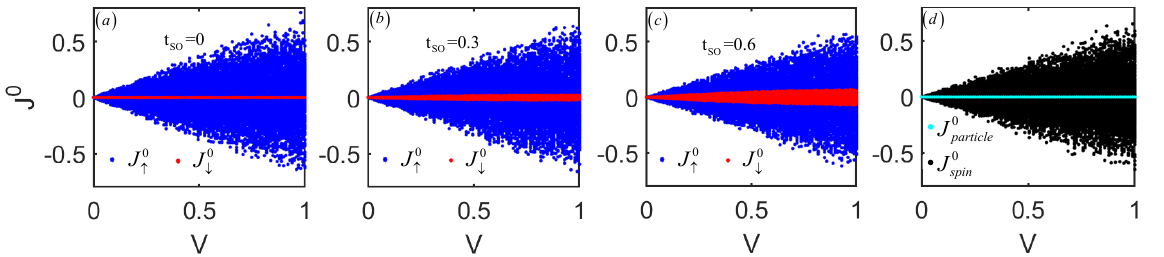}\\
	\caption{(a)-(c) The current $J_{\uparrow}^0$ and $J_{\downarrow}^0$ as a function of disorder strength $V$ with $V_{j\downarrow}=0$. (d) The particle and spin current as a function of disorder strength $V$ with $V_{j\uparrow}=V_{L+1-j\downarrow}$, the SOC strength $t_{SO}=0.3$. The temperature $T=0$, the chemical potentials $\mu_L=\mu_R=0.1$, and the system size $L=10$ are fixed.}\label{FigSSS2}
\end{figure}

Next, we consider a model incorporating with spin-orbit coupling (SOC):
\begin{equation}\label{R.9}
	H=-t_S\sum_{j\sigma}(c_{j+1\sigma}^{\dagger}c_{j\sigma}+h.c.) +t_{SO}\sum_{j}(c_{j\uparrow}^{\dagger}c_{j+1\downarrow}+c_{j\downarrow}^{\dagger}c_{j+1\uparrow}+h.c.) +\sum_{j\sigma}V_{j\sigma}c_{j\sigma}^{\dagger}c_{j\sigma},
\end{equation}
where $t_S$ is the hopping amplitude and $t_{SO}$ denotes the SOC strength. We first consider the case where only spin-up particles experience the disorder ($V_{j\downarrow}=0$, $V_{j\uparrow}\in[-V,V]$). 
When $t_{SO}=0$, spin-up and spin-down particles propagate independently. Consequently, the current $J^{0}_{\uparrow}\ne 0$, while $J^{0}_{\downarrow}=0$, due to the inversion symmetry, as illustrated in Fig. \ref{FigSSS2}(a). However, for $t_{SO}\ne 0$, the SOC enables spin-flip processes, leading to a finite 
$J^{0}_{\downarrow}$, as shown in Fig. \ref{FigSSS2}(b) and (c). The maximum spin-down current increases with $t_{SO}$, reflecting the SOC-assisted activation of the otherwise inert spin-down channel.

We then consider another type of spin-dependent disorder with inversion symmetry: $V_{j\uparrow}=V_{(L+1-j)\downarrow}$ (We note that our following discussion only requires the condition to be satisfied as long as the distributions of $V_{j\uparrow}$ and $V_{j\downarrow}$ themselves do not exhibit reflection symmetry, and it is not necessary for $V_j$ to be disordered). Under this configuration, the spin-up potential at site $j$ is mirrored by the spin-down potential at site $L+1-j$, and the hopping and SOC terms remain invariant under the combined operation of inversion and spin flip (the same phenomenon will occur even $t_{so}=0$). As a result, the spin-resolved currents satisfy $J_{\uparrow}^0=-J_{\downarrow}^0$. Therefore, the total particle current $J_{particle}^0=J_{\uparrow}^0+J_{\downarrow}^0$ vanishes, while the spin current $J_{spin}^0=\frac{1}{2}(J_{\uparrow}^0-J_{\downarrow}^0)$ is finite, which can be verified from numerical calculations as illustrated in Fig. \ref{FigSSS2}(d).

These findings demonstrate that the combination of spin-dependent disorder and SOC can be harnessed to realize controllable spin currents without net particle transport. Such spin-selective transport behavior has potential applications in the design of spin filters, spin valves, and other spintronic devices. In particular, the ability to generate pure spin currents through disorder engineering and SOC offers a promising route toward low-dissipation information processing. Our results also provide insights for implementing similar mechanisms in experimental platforms such as cold atoms, mesoscopic conductors, and synthetic quantum materials with tunable spin and dissipation control.

%we use a two-site model as an example to demonstrate the relationship between the sample average of the square of the disorder-induced current and the disorder strength $V$. The Hamiltonian of the model is$H_{S}=-(c_{1}^{\dagger}c_{2}+h.c.)+\sum_{j=1}^{2}V_jc_{j}^{\dagger}c_{j}$, where the potential $V_j$ is randomly distributed in $[-V, V]$. The on-site loss terms are $L_{j}=\sqrt{\gamma}c_{j}$, with $j=1, 2$, and exhibit inversion symmetry. The transmission function can be derived
%\begin{equation}\label{S.6.1}
%\tau_{0,l}(\omega)=\gamma\frac{\bar{V}_2^2-\bar{V}_1^2}{(\bar{V}_1\bar{V}_2-\bar{\gamma}^2-1)^2+\bar{\gamma}^2(\bar{V}_1+\bar{V}_2)^2},
%\end{equation}
%where the shifted potentials $\bar{V}_1=\omega-V_1$ and  $\bar{V}_2=\omega-V_2$ are randomly distributed in $[\omega-V,\omega+V]$, and $\tilde{\gamma}=\gamma+\frac{\Gamma}{2}$. In the weak disorder and weak dissipation limit, where $\bar{V},\bar{\gamma}\ll 1$, the current can be approximated by
%\begin{equation}\label{S.6.2}
%J^0=\frac{c}{2}(V_1-V_2),
%\end{equation}
%with $c=\gamma\Gamma\int\frac{d\omega}{h}f(\omega)\frac{4\omega} {4\omega^2(\gamma+\frac{\Gamma}{2})^2+[\omega^2-(\gamma+\frac{\Gamma}{2})^2-1]^2}$.
%From Eq. \ref{S.6.2} we see that the current vanishes $J^0=0$ if the potential is symmetric $V_1=V_2$. Conversely, the current reaches its maximum if the potential is the ``most'' asymmetric $V_1=-V_2=\pm V$. Hence the envelope curve of $J^0$ is given by $J^0\sim\pm cV$.

The disorder average of $(J^0)^2$ is
\begin{equation}\label{S.6.3a}
	\langle(J^0)^2\rangle=\frac{1}{4V^2}\int_{-V}^{V}\frac{c^2}{4}(V_1-V_2)^2dV_1dV_2=\frac{1}{6}c^2V^2.
\end{equation}
It can be shown numerically that this scaling relation $\langle(J^0)^2\rangle\sim V^2$ holds for large system size $N$ (see Fig. 2(d) in the main text).

\section{IX. Response current and the Wiedemann-Franz law}
In this section, we focus on the response current $\delta J^{\lambda}_1$. In the presence of chemical potential and temperature gradients, the change in currents can be written as
\begin{equation}\label{S.7.1}
	\left(
	\begin{array}{c}
		\delta J_1^{0} \\
		\delta J_1^{Q} \\
	\end{array}
	\right)=\left(
	\begin{array}{cc}
		L_{11} & L_{12} \\
		L_{21} & L_{22} \\
	\end{array}
	\right)\left(
	\begin{array}{c}
		\frac{\delta\mu}{T} \\
		\frac{\delta T}{T^2} \\
	\end{array}
	\right),
\end{equation}
where $\delta J_1^{Q}=\delta J_1^{1}-\mu_L\delta J_1^{0}-\delta\mu J_0^{0}$,  and $\mathbf{L}$ is the Onsager matrix \cite{ButcherPN,Chiaracane}. Then, the electrical conductance $G$, the thermal conductance $K$, and the Seebeck factor (or thermopower) $S$ can be expressed as 
\begin{equation}\label{S.7.2}
	G=\frac{e^2}{T}L_{11},~~~~~~K=\frac{1}{T^2}\frac{\det|\mathbf{L}|}{L_{11}},~~~~~~S=\frac{1}{eT}\frac{L_{12}}{L_{11}}.
\end{equation}

In the linear response regime, the Fermi function difference is
\begin{equation}\label{S.7.3}
	\delta f=f_L-f_R=\frac{\partial f}{\partial T}\delta T+\frac{\partial f}{\partial\mu}\delta\mu =-f^{'}(\omega)\bigg[\frac{\omega-\mu}{T}\delta T+\delta\mu\bigg],
\end{equation}
with $f^{'}(\omega)=\frac{-1}{4k_{B}T\cosh^2(\frac{\omega-\mu}{2k_{B}T})}$. In the zero-temperature limit, $f^{'}(\omega)\rightarrow -\delta(\omega-\mu)$, and in the high-temperature limit, $f^{'}(\omega)\rightarrow\frac{-1}{4k_B T}$. It is convenient to define the integral
\begin{equation}\label{S.7.4}
	I_k=-\int d\omega(\omega-\mu)^{k}f^{'}(\omega)\tau_1(\omega),
\end{equation}
from which we find the Onsager matrix elements are $L_{11}=\frac{T}{h}I_0$, $L_{12}=L_{21}=\frac{T}{h}I_1$, $L_{22}=\frac{T}{h}I_2$. Hence, all the response functions in Eq. (\ref{S.7.2}) can be expressed in terms of $I_k$. If $\tau_1(\omega)$ is a continuous function near the chemical potential $\mu$, then in the low-temperature limit, Eq. (\ref{S.7.4}) can be approximated by
\begin{equation}\label{S.7.5}
	I_0=\tau_1(\mu),~~~~~~I_1=\frac{\pi^2}{3}(k_BT)^2\frac{d\tau_1(\mu)}{d\mu},~~~~~~I_2=\frac{\pi^2}{3}(k_BT)^2\tau_1(\mu).
\end{equation}
Then the ratio $\frac{K}{G}$ is given by
\begin{equation} 
	\frac{K}{G}=\frac{1}{e^2T}\frac{I_0I_2-I_1^2}{I_0^2}=\mathcal{L}T[1-\frac{\pi^2}{3}(k_BT)^2(\frac{\tau_1^{\prime}(\mu)}{\tau_1(\mu)})^2], 
\end{equation}
where the Lorenz number is  $\mathcal{L}=\frac{\pi^2}{3}(\frac{k_B}{e})^2$. The Wiedemann-Franz law holds if $|\frac{\tau_1^{\prime}(\mu)}{\tau_1(\mu)}|\ll(k_BT)^{-1}$. In the absence of gain and loss, $|\frac{\tau_1^{\prime}(\mu)}{\tau_1(\mu)}|$ becomes very large near the band edge or outside the band, leading to a breakdown of the Wiedemann-Franz law. However, gain and loss smooth the transmission function $\tau_1(\omega)$, ensuring that the Wiedemann-Franz law holds across all energy regimes.

\section{X. $J_0^0$ variation with size in the presence of skin effect}
We consider the tight-binding Hamiltonian $H_{S}=-t_S\sum_{j=1}^{N-1}(c_{j}^{\dagger}c_{j+1}+h.c.)$ with bond loss
$L_{j\in[2,N]}=\sqrt{\gamma_{-}}(c_j-ic_{j-1})$. 
Under open boundary conditions, we introduce two loss channels: $L_{1}=\sqrt{\gamma_{-}}c_{1}$ and $L_{N+1}=\sqrt{\gamma_{-}}c_{N}$ at the boundaries. All  eigenstates are localized near the left end for $t_S>0$ and near the right end for $t_S<0$. We assume  $(\tilde{\Sigma}_{L})_{11}=(\tilde{\Sigma}_{R})_{NN}=-\frac{i\Gamma}{2}$. The transmission function can be expressed as
\begin{equation}\label{S.8.2}
	\tau_{0,l}(\omega)=2\gamma_{-}\Gamma\sum_{j=1}^{N}[|G_{1,j}^{\mathcal{R}}|^2-|G_{N,j}^{\mathcal{R}}|^2] +2\gamma_{-}\Gamma\sum_{j=1}^{N-1}Im[G_{1,j}^{\mathcal{R}}(G_{1,j+1}^{\mathcal{R}})^{*}-G_{N,j}^{\mathcal{R}}(G_{N,j+1}^{\mathcal{R}})^{*}].
\end{equation}
Here we have omitted the subsrcipt $S$ in the Green's functions $\mathbf{G}_S^{\mathcal{R}}$ for convenience, and $G_{i,j}^{\mathcal{R}}$ denotes the $i$-th row and 
$j$-th column element of $\mathbf{G}_S^{\mathcal{R}}$.

To derive the expression for the transmission function, the main task is to calculate the matrix elements $G_{1,j}^{\mathcal{R}}$ and $G_{N,j}^{\mathcal{R}}$. Noting that the inverse Green's function $(\mathbf{G}_S^{\mathcal{R}})^{-1}=\omega-\mathbf{X}-\tilde{\bm{\Sigma}}_{L}^{R}-\tilde{\bm{\Sigma}}_{R}^{R}$ is a tridiagonal matrix:
\begin{equation}\label{S.8.3}
	(\mathbf{G}_S^{\mathcal{R}})^{-1}=
	\left(
	\begin{array}{cccccc}
		\omega-X_{1,1}-(\tilde{\bm{\Sigma}}_{L}^{\mathcal{R}})_{11} & -X_{1,2} & ~ & ~ & ~ & ~ \\
		-X_{2,1} & \omega-X_{2,2} & -X_{2,3} & ~ & ~ & ~ \\
		~ & -X_{3,2} & \omega-X_{3,3} & -X_{3,4} & ~ & ~ \\
		~ & ~ & \ddots & \ddots & \ddots & ~ \\
		~ & ~ & ~ & -X_{N-1,N-2} & \omega-X_{N-1,N-1} & -X_{N-1,N} \\
		~ & ~ & ~ & ~ & -X_{N,N-1} & \omega-X_{N,N}-(\tilde{\bm{\Sigma}}_{R}^{\mathcal{R}})_{NN} \\
	\end{array}
	\right),
\end{equation}
where $X_{i,j}$ is the ($i,j$)-th element of the damping matrix $\mathbf{X}$, $(\tilde{\bm{\Sigma}}_{L}^{\mathcal{R}})_{11}$ is
the ($1, 1$)-th element of  $\tilde{\bm{\Sigma}}_{L}^{\mathcal{R}}$, and $(\tilde{\bm{\Sigma}}_{R}^{\mathcal{R}})_{NN}$ is
the ($N, N$)-th element of  $\tilde{\bm{\Sigma}}_{R}^{\mathcal{R}}$. 
The matrix elements of Green's functions $\mathbf{G}_S^{\mathcal{R}}$ can be represented by \cite{Lavis}
\begin{equation}\label{S.8.4}
	G_{i,j}^{\mathcal{R}}=
	\begin{cases}
		X_{i,i+1}X_{i+1,i+2}\cdots X_{j-1,j}\frac{\Delta_{1,i-1}\Delta_{j+1,N}}{\Delta_{1,N}},~~~~~~i<j, \\ \frac{\Delta_{1,i-1}\Delta_{j+1,N}}{\Delta_{1,N}},~~~~~~~~~~~~~~~~~~~~~~~~~~~~~~~~~~~~~~~i=j, \\ X_{j+1,j}X_{j+2,j+1}\cdots X_{i,i-1}\frac{\Delta_{1,j-1}\Delta_{i+1,N}}{\Delta_{1,N}},~~~~~~i>j,
	\end{cases}
\end{equation}
where $\Delta_{ij}$ is the determinant of the submatrix of $(\mathbf{G}_S^{\mathcal{R}})^{-1}$ from $i$-th row and $i$-th column to $j$-th row and $j$-th column \cite{Saha}, which is determined by the iterative equations:
\begin{equation}\label{S.8.5}
	\begin{split}
		\left(
		\begin{array}{c}
			\Delta_{1,j} \\
			\Delta_{1,j-1} \\
		\end{array}
		\right)=\left(
		\begin{array}{cc}
			\omega-\tilde{X}_{j,j} & -X_{j-1,j}X_{j,j-1} \\
			1 & 0 \\
		\end{array}
		\right)\left(
		\begin{array}{c}
			\Delta_{1,j-1} \\
			\Delta_{1,j-2} \\
		\end{array}
		\right) \\
		\left(
		\begin{array}{c}
			\Delta_{j,N} \\
			\Delta_{j+1,N} \\
		\end{array}
		\right)=\left(
		\begin{array}{cc}
			\omega-\tilde{X}_{j,j} & -X_{j,j+1}X_{j+1,j} \\
			1 & 0 \\
		\end{array}
		\right)\left(
		\begin{array}{c}
			\Delta_{j+1,N} \\
			\Delta_{j+2,N} \\
		\end{array}
		\right)
	\end{split}
\end{equation}
where $\Delta_{1,0}=1$, $\Delta_{1,1}=\omega-\tilde{X}_{1,1}$, $\Delta_{N,N}=\omega-\tilde{X}_{N,N}$ and $\Delta_{N+1,N}=1$, and $\tilde{X}_{1,1}=X_{1,1}+(\tilde{\bm{\Sigma}}_{L}^{\mathcal{R}})_{11}$, $\tilde{X}_{j,j}=X_{j,j}$ for $j=2,3,\cdots,N-1$, $X_{N,N}=X_{N,N}+(\tilde{\bm{\Sigma}}_{R}^{\mathcal{R}})_{NN}$. Defining the transfer matrix
\begin{equation}\label{S.8.6}
	\begin{split}
		\bar{T}_j=\left(
		\begin{array}{cc}
			\omega-X_{jj} & -X_{j-1,j}X_{j,j-1} \\
			1 & 0 \\
		\end{array}
		\right), ~~~~~~~~
		\tilde{T}_j=\left(
		\begin{array}{cc}
			\omega-X_{jj} & -X_{j,j+1}X_{j+1,j} \\
			1 & 0 \\
		\end{array}
		\right).
	\end{split}
\end{equation}
Eq. (\ref{S.8.6}) can be rewritten as
\begin{equation}\label{S.8.7}
	\begin{split}
		& \begin{cases}
			\left(
			\begin{array}{c}
				\Delta_{1,j} \\
				\Delta_{1,j-1} \\
			\end{array}
			\right)=\bar{T}_{j}\bar{T}_{j-1}\cdots\bar{T}_1\left(
			\begin{array}{c}
				1 \\
				\frac{(\tilde{\bm{\Sigma}}_{L}^{\mathcal{R}})_{11}}{X_{0,1}X_{1,0}} \\
			\end{array}
			\right),~~~~~~~~~~~~~~~~~~~~~~~~~~~~1\leq j<N, \\
			\left(
			\begin{array}{c}
				\Delta_{1,N} \\
				\Delta_{1,N-1} \\
			\end{array}
			\right)=\left(
			\begin{array}{cc}
				1 & -(\tilde{\bm{\Sigma}}_{R}^{\mathcal{R}})_{NN} \\
				0 & 1 \\
			\end{array}
			\right)\bar{T}_{N}\bar{T}_{N-1}\cdots\bar{T}_1\left(
			\begin{array}{c}
				1 \\
				\frac{(\tilde{\bm{\Sigma}}_{L}^{\mathcal{R}})_{11}}{X_{0,1}X_{1,0}} \\
			\end{array}
			\right),~~~~~~j=N.
		\end{cases} \\
		& \begin{cases}
			\left(
			\begin{array}{c}
				\Delta_{1,N} \\
				\Delta_{2,N} \\
			\end{array}
			\right)=\left(
			\begin{array}{cc}
				1 & -(\tilde{\bm{\Sigma}}_{L}^{\mathcal{R}})_{11} \\
				0 & 1 \\
			\end{array}
			\right)\tilde{T}_{1}\tilde{T}_{2}\cdots\tilde{T}_N\left(
			\begin{array}{c}
				1 \\
				\frac{(\tilde{\bm{\Sigma}}_{R}^{\mathcal{R}})_{NN}}{X_{N,N+1}X_{N+1,N}} \\
			\end{array}
			\right),~~~~~~~j=1, \\
			\left(
			\begin{array}{c}
				\Delta_{j,N} \\
				\Delta_{j+1,N} \\
			\end{array}
			\right)=\tilde{T}_{j}\tilde{T}_{j+1}\cdots\tilde{T}_N\left(
			\begin{array}{c}
				1 \\
				\frac{(\tilde{\bm{\Sigma}}_{R}^{\mathcal{R}})_{NN}}{X_{N,N+1}X_{N+1,N}} \\
			\end{array}
			\right),~~~~~~~~~~~~~~~~~1<j\leq N.
		\end{cases}
	\end{split}
\end{equation}
Here, for convenience, we define $X_{N,N+1}=X_{N,1}$, $X_{N+1,N}=X_{1,N}$, $X_{1,0}=X_{1,N}$ and $X_{0,1}=X_{N,1}$.

Substituting Eq. (\ref{S.8.4}) into Eq. (\ref{S.8.2}), we obtain
\begin{equation}\label{S.8.8}
	\begin{split} 
		\tau_{0,l}(\omega)=& 2\gamma_{-}\Gamma\sum_{j=1}^{N}[(-t_S-\gamma_{-})^{2N-2j}-(-t_S+\gamma_{-})^{2N-2j}]\frac{|\Delta_{1,j-1}|^2}{|\Delta_{1,N}|^2} \\ +&2\gamma_{-}\Gamma\sum_{j=1}^{N-1}[(-t_S-\gamma_{-})^{2N-2j-1}+(-t_S+\gamma_{-})^{2N-2j-1}]Im\frac{\Delta_{1,j}\Delta_{1,j-1}^{*}}{|\Delta_{1,N}|^2}.
	\end{split}
\end{equation}
The transfer matrix (\ref{S.8.6}) becomes
\begin{equation}\label{S.8.9}
	\bar{T}=\tilde{T}=
	\left(
	\begin{array}{cc}
		\omega+2i\gamma_{-} & \gamma_{-}^2-t_S^2 \\
		1 & 0 \\
	\end{array}
	\right).
\end{equation}
The eigenvalues are $\lambda_{\pm}=\frac{\omega+2i\gamma_{-}\pm\Delta}{2}$, with $\Delta=\sqrt{\omega^2-4t^2+4i\omega\gamma_{-}}$. From Eq. (\ref{S.8.7}), we get
\begin{equation}\label{S.8.10}
	\begin{split}
		& \Delta_{1,j}=\frac{1}{\Delta}(s_{j+1}+\frac{i\Gamma}{2}s_{j}),~~~~~~1\leq j<N, \\ & \Delta_{1,N}=\frac{1}{\Delta}[s_{N+1}+i\Gamma s_{N}-\frac{\Gamma^2}{4}s_{N-1}]~~~~~~j=N,
	\end{split}
\end{equation}
where $s_j=\lambda_{+}^j-\lambda_{-}^j$.

We consider the special case $t_S=-\gamma_{-}$ and the low-energy limit $\omega\ll|t_S|, \gamma_{-}$. Up to leading order, $\lambda_{+}=0$ and $\lambda_{-}=-(\omega+2it)$, and the transmission function in Eq. (\ref{S.8.8}) can be simplified to
\begin{equation}\label{S.8.11}
	\tau_{0,l}(\omega)=\frac{4\gamma_{-}\Gamma}{|\Delta_{1,N}|^2}\sum_{j=1}^{N-1}(-2t_S)^{2N-2j-1} Re\bigg[\Delta_{1,j-1}(t_S\Delta_{1,j-1}^{*}+\frac{i}{2}\Delta_{1,j}^{*})\bigg],
\end{equation}
where
\begin{equation}\label{S.8.12}
	\begin{split}
		& \Delta_{1,0}(t\Delta_{1,0}^{*}+\frac{i}{2}\Delta_{1,1}^{*})=\frac{\Gamma}{4}+\frac{i\omega}{2}, \\
		& \Delta_{1,j-1}(t\Delta_{1,j-1}^{*}+\frac{i}{2}\Delta_{1,j}^{*}) =\frac{i\omega}{2}[\omega^2+\frac{1}{4}(4t_S-\Gamma)^2]\times(\omega^2+4t_S^2)^{j-2}, \\ & |\Delta_{1,N}|^2=[\omega^2+\frac{1}{4}(4t_S-\Gamma)^2]^2\times(\omega^2+4t_S^2)^{N-2}.
	\end{split}
\end{equation}
Substituting Eq. (\ref{S.8.12}) into Eq. (\ref{S.8.11}), we obtain
\begin{equation}\label{S.8.13}
	\tau_{0,l}(\omega)=\frac{2\gamma_{-}^2\Gamma^2}{[\omega^2+\frac{1}{4}(4t_S-\Gamma)^2]^2}(1+\frac{\omega^2}{4t_S^2})^{2-N} \sim\frac{2e^2\gamma_{-}^2\Gamma^2}{[\omega^2+\frac{1}{4}(4t_S-\Gamma)^2]^2}e^{-\frac{\omega^2}{4t_S^2}N}.
\end{equation}
Similarly, for $t_S=\gamma_{-}$, we can derive
\begin{equation}\label{S.8.14}
	\tau_{0,l}(\omega)\sim-\frac{2e^2\gamma_{-}^2\Gamma^2}{[\omega^2+\frac{1}{4}(4t-\Gamma)^2]^2}e^{-\frac{\omega^2}{4t_S^2}N}.
\end{equation}
Though Eq. (\ref{S.8.13}) and Eq. (\ref{S.8.14}) are derived for $|t_S|=|\gamma_{-}|$, the scaling $\tau_{0,l}\sim e^{-\omega^2N}$ provides a good approximation for $|t_S|\geq|\gamma_{-}|$ based on numerical calculations.
Then the current at zero temperature is
\begin{equation}\label{S.8.15}
	J_0^0=\int_{-\infty}^{\mu}\frac{d\omega}{h}\tau_{0,l}(\omega)\sim\sqrt{\frac{\pi}{4hN}}[1+erf(\mu\sqrt{N})],
\end{equation}
where $erf(x)=\frac{2}{\sqrt{\pi}}\int_{0}^{x}e^{-t^2}dt$ is the error function. The asymptotic expansion of error function is $erf(x)\sim1-\frac{e^{-x^2}}{\sqrt{\pi}x}$ as  $x\rightarrow\infty$, and $erf(x)\sim-1-\frac{e^{-x^2}}{\sqrt{\pi}x}$ as $x\rightarrow-\infty$. Hence, in the thermodynamic limit $N\rightarrow\infty$, we obtain $J_0^0\sim N^{-\frac{1}{2}}$ for $\mu\geq0$, and $J_0^0\sim \frac{e^{-N}}{N}$ for $\mu<0$. Therefore, the current corresponding to the skin effect decreases as the size increases and vanishes as the size tends to infinity.

%%%%%%%%%%%%%%%%%%%%%%%%%%%%%%%%%%%%%%%%%%%%%


\begin{thebibliography}{36}
\bibitem{explainOLB} ``Original coherent formulation'' refers to the standard Landauer-B\"{u}ttiker transport setup without dissipation; this excludes the later voltage-probe extension~\cite{ButtikerVP}, where decoherence is introduced phenomenologically via phase-randomizing probes (see End Matter).
\bibitem{explain2} It is easy to prove that when the Hamiltonian $\mathbf{H}$, the loss matrix $\mathbf{P}$, and the gain matrix $\mathbf{Q}$ satisfy inversion symmetry, $\mathbf{G}_{S}^{\mathcal{R}}$ and $\mathbf{G}_{S}^{\mathcal{A}}$
	also satisfy inversion symmetry. Therefore, $\mathbf{O}$ also corresponds to $\mathbf{G}_{S}^{\mathcal{R}}$ and $\mathbf{G}_{S}^{\mathcal{A}}$.
\bibitem{explainF} This conclusion is generally valid; however, as we will discuss later, exceptions arise when the system exhibits non-Hermitian skin effects.
\end{thebibliography}
\end{document}